\documentclass[number, preprint, sort&compress, 1p]{elsarticle}
\pdfoutput=1
\usepackage[utf8]{inputenc}
\usepackage[T1]{fontenc}

\usepackage{graphicx}
\usepackage[english, american, ngerman]{babel}
\usepackage{lmodern}
\usepackage{subcaption}
\usepackage{flushend}
\usepackage{xcolor}
\usepackage{relsize}
\usepackage{mathtools}
\usepackage{wrapfig}
\usepackage{xcolor}
\usepackage{colortbl}
\usepackage{longtable}
\usepackage{relsize}
\usepackage{pgfplots}
\usepackage{dsfont}
\usepackage{bibentry}
\usepackage{hyperref}
\usepackage{amsmath}
\usepackage{amssymb}

\let \vec \mathbf

\definecolor{cd1}{RGB}{62, 68, 76}
\definecolor{cd2}{RGB}{0, 190, 255}
\definecolor{cd3}{RGB}{0, 81, 158}
\definecolor{cd4}{RGB}{159, 153, 152}
\definecolor{cd5}{RGB}{255, 213, 0}

\definecolor{mycolor1}{rgb}{0.2471  0.6275  1.0000}
\definecolor{mycolor2}{rgb}{1,0.1882,0.1882}
\definecolor{mycolor3}{rgb}{0.6235,0.6000,0.5961}
\definecolor{mycolor4}{rgb}{0.9020,0.9020,0.9020}
\definecolor{mycolor5}{rgb}{0.24314,0.26667,0.29804}
\definecolor{mycolor6}{rgb}{0, 0.3176, 0.6196}

\pgfplotsset{compat=newest}
\pgfplotsset{plot coordinates/math parser=false}
\newlength\figureheight
\newlength\figurewidth 

\begin{document}

\selectlanguage{american}

\begin{frontmatter}


\title{A Port-Hamiltonian Approach to Modeling the Structural Dynamics of Complex Systems\tnoteref{copyright}\tnoteref{doi}}
\tnotetext[copyright]{\copyright~2020. This manuscript version is made available under the CC-BY-NC-ND 4.0 license \url{http://creativecommons.org/licenses/by-nc-nd/4.0/}}   
\tnotetext[doi]{Published version: \url{https://doi.org/10.1016/j.apm.2020.07.038}}
\author[isys]{Alexander Warsewa\corref{cor}}
\ead{warsewa@isys.uni-stuttgart.de}
\author[isys]{Michael B{\"o}hm}
\author[isys]{Oliver Sawodny}
\author[isys]{Cristina Tar{\'i}n}
\cortext[cor]{Corresponding author.}
\address[isys]{Institute for System Dynamics, University of Stuttgart, Waldburgstraße 17/19, 70563 Stuttgart, Germany}

\begin{abstract}
With this contribution, we give a complete and comprehensive framework for modeling the dynamics of complex mechanical structures as port-Hamiltonian systems. This is motivated by research on the potential of lightweight construction using active load-bearing elements integrated into the structure. Such adaptive structures are of high complexity and very heterogeneous in nature. Port-Hamiltonian systems theory provides a promising approach for their modeling and control. Subsystem dynamics can be formulated in a domain-independent way and interconnected by means of power flows. The modular approach is also suitable for robust decentralized control schemes. 

Starting from a distributed-parameter port-Hamiltonian formulation of beam dynamics, we show the application of an existing structure-preserving mixed finite element method to arrive at finite-dimensional approximations. In contrast to the modeling of single bodies with a single boundary, we consider complex structures composed of many simple elements interconnected at the boundary. This is analogous to the usual way of modeling civil engineering structures which has not been transferred to port-Hamiltonian systems before. A block diagram representation of the interconnected systems is used to generate coupling constraints which leads to differential algebraic equations of index one. After the elimination of algebraic constraints, systems in input-state-output (ISO) port-Hamiltonian form are obtained. Port-Hamiltonian system models for the considered class of systems can also be constructed from the mass and stiffness matrices obtained via conventional finite element methods. We show how this relates to the presented approach and discuss the differences, promoting a better understanding across engineering disciplines. A Matlab framework is available on \url{http://github.com/awarsewa/ph_fem/} to facilitate the application of the methods to different problems.
\end{abstract}



\end{frontmatter}

\section{Introduction}
Lightweight structures have become a reality for mass-sensitive applications, such as large civil engineering structures and many more. For passive structures, these designs present in most cases a minimum in terms of required mass under given safety limitations and user comfort constraints. However, it is possible to stay within these limits while further reducing the total embodied mass significantly by introducing \emph{adaptive structures}, which is referred to as \emph{ultralightweight design}. Through their various actuators, these structures can react and adapt to external loads and disturbances~--~both static and dynamic~--~in order to minimize element stress and at the same time maximize lifetime expectancy~\cite{Korkmaz2011}.

In light of worldwide population growth, demographic aging and ever increasing standard-of-living, construction activities are at an all-time high and are expected to increase further drastically within the next 20 to 30 years. This in mind, the technology of active and adaptive structures can help save millions of tons of concrete and steel and significantly reduce waste production and $\mathrm{CO}_2$-emissions of the construction industry~\cite{Sobek2014}. 

Nevertheless, methods for analysis and control of adaptive structures need to be developed in order to make this technology readily available. This requires accurate mathematical models of these structures. The current state of the art for modeling as well as static and dynamic analysis of civil engineering structures is finite element modeling (FEM). However, most often the elements are connected via their nodal displacements, not by the energy flow from one to another. This technique impedes the integration of actuators into the dynamic model of a structure. At the same time, it can be difficult for these structures to design decentralized control schemes with guaranteed asymptotic stability.

In this context, port-Hamiltonian systems have the potential to provide a framework for modeling both passive and active structures and for seamless and straightforward integration of actuators from different physical domains, such as electric motors or hydraulic cylinders. Furthermore, robust decentralized control strategies based on passivity based control can naturally be used for these structures.

Truss structures and frames can be modeled as an interconnection of one-dimensional linear elastic beam elements and rods.~Although our use case is adaptive structures, the structural dynamics of other complex systems, such as in aerospace, can also be described this way. For simulation and control purposes, it is expedient to describe the system dynamics in state space. In the case of port-Hamiltonian systems, a common state space representation is the input-state-output (ISO) form \cite{van2014port}
	\begin{align}\label{eq:ISO_form}
			\begin{split}
				\dot{\vec x} &= \left(\vec J - \vec R\right) \frac{\mathrm{d}H}{\mathrm{d} \vec x} + \vec G \vec u, \\
				\vec y &= \vec G^\mathrm{T} \frac{\mathrm{d}H}{\mathrm{d} \vec x}
		\end{split}
	\end{align}
with the state vector $\vec x \in \mathbb R^n$ which contains the energy variables of the physical system, the skew-symmetric interconnection matrix $\vec J \in \mathbb R^{n\times n}$, the positive semi-definite symmetric dissipation matrix $\vec R \in \mathbb R^{n\times n}$ and the output matrix $\vec G \in \mathbb R^{n\times n_u}$. Inputs and outputs are collocated and $n_u$ is the number of both inputs and outputs. Evaluation of the Hamiltonian $H(\vec x)$ yields the energy stored in the system. For linear port-Hamiltonian systems, the Hamiltonian is quadratic and can be expressed as $H(\vec x) = \frac{1}{2} \vec x^\mathrm{T} \vec Q \vec x$, where $\vec Q \in \mathbb R^{n\times n}$ is a symmetric positive definite matrix. Flow and effort variables are defined as $\vec f := \dot{\vec x}$ and $\vec e := \mathrm{d} H/\mathrm{d} \vec x$, respectively, with the power product $\vec e^\mathrm{T} \vec f \leq 0$.

The dynamics of complex linear elastic systems obtained from FE models can be stated as second-order differential equations of the following form
\begin{equation}\label{eq:mech_2nd_order}
	\vec M_\text{fe} \ddot{\vec q} + \vec D_\text{fe} \dot{\vec q} + \vec K_\text{fe} \vec q = \vec{f}_\text{ext},
\end{equation}
where $\vec M_\text{fe} \in \mathbb R^{n_\text{DOF}\times n_\text{DOF}}$ is the mass matrix, $\vec D_\text{fe} \in \mathbb R^{n_\text{DOF}\times n_\text{DOF}}$ the damping matrix and $\vec K_\text{fe} \in \mathbb R^{n_\text{DOF}\times n_\text{DOF}}$ the stiffness matrix. The system's global degrees of freedom (DOFs) are collected in the vector $\vec q \in \mathbb R^{n_\text{DOF}}$ and the external forces acting on the structure in $\vec f_\text{ext} \in \mathbb R^{n_{DOF}}$. Any such system can be written in ISO port-Hamiltonian form~\eqref{eq:ISO_form} as long as $\vec M_\text{fe}$ and $\vec K_\text{fe}$ are symmetric positive definite and $\vec D_\text{fe}$ is symmetric positive semi-definite, which is usually the case. By choosing the state vector as $\vec x = \begin{bmatrix} \vec M_\text{fe} \dot{\vec q} & \vec q \end{bmatrix}$, we obtain
\begin{align}\label{eq:ph_iso_mech}
\begin{split}
	\dot{\vec x} &= \Bigg(\underbrace{\begin{bmatrix} 0 & -\vec I \\ \vec I & 0 \end{bmatrix}}_{\vec J} - \underbrace{\begin{bmatrix} \vec D_\text{fe} & 0\\ 0 & 0 \end{bmatrix}}_{\vec R} \Bigg)\underbrace{\begin{bmatrix} \vec M_\text{fe}^{-1} & 0 \\ 0 & \vec K_\text{fe} \end{bmatrix}}_\vec Q \vec x + \underbrace{\begin{bmatrix} \vec I \\ 0 \end{bmatrix}}_\vec G \vec f_\text{ext}, \\
	\vec y &= \vec G^\text T \vec Q \vec x.
\end{split}
\end{align}
The concentrated parameter equations \eqref{eq:mech_2nd_order} are approximations of the dynamics of elastic mechanical systems that result from the spatial discretization of infinite-dimensional component models and their interconnection. In conventional FEM, this is done by applying the principle of virtual work and the Galerkin method. By choosing polynomial approximation bases for the DOFs, element mass and stiffness matrices directly result from the integration of the potential and kinetic energy terms in the variational formulation of the system Hamiltonian. Transforming element DOFs to global coordinates and adding terms corresponding to the same DOFs, the system mass and stiffness matrices and \eqref{eq:mech_2nd_order} or \eqref{eq:ph_iso_mech} are obtained.

If we were only interested in a port-Hamiltonian formulation of linear elastic structures, there would be no need to proceed any further. However, we are interested the interconnection of such systems with elements from non-mechanical or hybrid domains. Suppose, that such elements are described by partial-differential equations (PDEs) or include nonlinear relations. In those cases, it is not clear how to obtain a port-Hamiltonian formulation of the coupled system, given only the mechanical part \eqref{eq:mech_2nd_order}. This is why we present the port-Hamiltonian way of modeling linear elastic structures using suitable FE approaches and a method for the coupling of basic elements. In each step -- from the infinite-dimensional equations via spatial discretization to the interconnected concentrated parameter system -- the port-Hamiltonian structure is preserved.  From a control engineer's perspective, the passivity property of the port-Hamiltonian system representation can be exploited for control design. Especially for nonlinear systems it eases stability proofs or finding suitable Lyapunov functions. Early lumping of infinite-dimensional systems with boundary controllers is also possible.

The port-Hamiltonian approach does not directly produce element mass and stiffness matrices or equations of the form \eqref{eq:ph_iso_mech}. Even though the procedure is significantly different to established FE methods, we show that the resulting systems are identical and can be transformed to \eqref{eq:ph_iso_mech} when the same approximation bases are used. To the best of our knowledge, such a connection has not been established before. With a thorough understanding of the port-Hamiltonian way of modeling linear elastic mechanical systems and its relation to more established FE methods, extensions are achieved more easily and mutual understanding across disciplines is promoted. In addition to presenting a comprehensive method for systems composed of beams -- which is the main focus of this contribution -- coupling with the hydraulic domain is shown in an example. This emphasizes that it is straightforward to construct models of complex multi-domain systems within the presented port-Hamiltonian framework.

\subsection{Related Work}
The dynamics of one-dimensional beam elements are by default described by PDEs. Physical systems of this type can be formulated as infinite-dimensional port-Hamiltonian systems without difficulties. For the corresponding theory on distributed-parameter port-Hamiltonian systems, see e.\,g. \cite{zwart2009distributed, van2002hamiltonian}. In order to obtain systems of the form \eqref{eq:ISO_form}, a finite-dimensional approximation is necessary. A wide range of methods for the numerical approximation of PDEs is available \cite{quarteroni2008numerical}. As stated above, the finite element method is a common technique in structural dynamics. For a comprehensive treatise of the approach, see e.\,g. \cite{schwertassek2017dynamik} or \cite{hughes2012finite}. Discretization of infinite dimensional port-Hamiltonian systems, however, can not be done in the usual ways. Both the duality and the passivity property need to be preserved in the finite-dimensional approximation. The former can be achieved by using mixed FEM approaches. First methods for the structure-preserving spatial discretization of port-Hamiltonian systems were presented in \cite{talasila2002wave} and \cite{golo2004hamiltonian}. They use differential forms for the numerical approximation of one- and two-dimensional systems, where different approximation spaces are used for flows and efforts. Their method was extended to allow for higher-order spatial derivatives by Bassi et. al \cite{bassi2007algorithm} and for its application to a class of irreversible thermodynamic systems by Baaiu et al. \cite{baaiu2009structure}. For an accurate approximation, a relatively large number of finite elements is usually required with these approaches. Moulla et al. \cite{moulla2012pseudo} adopted a collocation method to discretize 1D port-Hamiltonian systems. Using polynomial bases, this results in higher accuracy for lower-order approximations. In all of the approaches mentioned up to this point, the discretization happens in the strong form of the balance equations. This results in rather strict compatibility conditions. Also, an input feed-through term is present in the finite-dimensional approximation which potentially complicates the application of control engineering methods. 

Cardoso-Ribeiro et al. \cite{cardoso2016piezoelectric} relax the compatibility conditions by using the method of \cite{moulla2012pseudo} in the weak formulation in order to apply it to model the dynamics of a beam excited by a piezoelectric patch. Mixed finite element discretization in the weak form was introduced in a general way by Kotyczka et al. \cite{kotyczka2018weak}. Their method can be applied to systems of any spatial dimension and resolves several issues that arise when approximating directly in the strong form. It requires an additional projection to ensure non-degeneracy of the power product which introduces additional degrees of freedom that can be used to tune the discretized models. The boundary ports of resulting models are of alternating causality which is not always desired. In this contribution we use the partitioned finite-element method (PFEM) of Cardoso-Ribeiro et al. \cite{cardoso2018structure}. It is closely related to the mixed Galerkin method used in \cite{kotyczka2018weak}. Integration by parts is, however, only applied to one of the dual balance equations. This way, the port-Hamiltonian structure is obtained in a direct fashion and no further projection is required. Moreover, no feed-through term is produced in the discretization process. Causality of the boundary ports depends on which balance equation is selected for partial integration. It does not alternate for individual elements. The method was also shown to be compatible with existing finite element software. PFEM is applied to the Mindlin plate in~\cite{brugnoli2019mindlin} and to the Kirchhoff-Love plate model in~\cite{brugnoli2019kirchhoff}. A symplectic Hamiltonian formulation of thin plates was also introduced by Li~et~al. in~\cite{li2016new} to obtain analytic solutions for the free vibration problem. It has recently been extended to account for the buckling problem in \cite{li2018buckling} and thin cylindrical shell structures in \cite{li2019hamiltonian}. While it constitutes a powerful method to obtain analytic solutions of boundary-value problems, it doesn't esasily extend to systems interconnected in a complex way. All in all, PFEM is considered the most suitable approach for the application described above. 

In available examples, the system usually consist of a single body and its boundary. Our use case differs in that we consider complex structures formed by an interconnection of many simple bodies at the boundary. This is a common approach in the modeling of civil engineering structures which we apply within the port-Hamiltonian framework in the following. To arrive at an automated way of generating models of arbitrary structures composed of beam elements, we take a further look at the coupling of elements. 

\subsection{Outline}
The structure of this contribution is as follows. In Sec.\,\ref{sec:truss_elements}, distributed-parameter port-Hamiltonian models are presented for the load cases of a linear beam element, i.\,e. bending, axial loads and torsion. Regarding bending, both Euler-Bernoulli and the Timoshenko bending theory are considered. Depending on the nature of the problem examined, the formulation with higher validity can be chosen. Given the basic component dynamics in PDE-form, the discretization process using PFEM is explained in Sec.\,\ref{sec:discretization}. Instead of performing the discretization process on each of the systems in Sec.\,\ref{sec:truss_elements} individually, it is stated on a more general system class level. The interconnection of subsystems is illustrated in Sec.\,\ref{sec:assembly}. Based on a mechanical network diagram analogy, an algorithm for the generation of coupling constraints is derived. 
After introducing and algebraically manipulating coupling constraints, the interconnected systems can be written as differential-algebraic equations (DAEs) of index one in ISO port-Hamiltonian form. Reduction to ordinary differential equations (ODEs) on a constrained state space is outlined thereupon. Then, it is shown that the systems can be brought to the form \eqref{eq:ph_iso_mech} using simple transformations. In Sec.\,\ref{sec:example}, a range of examples on the use of the approach is presented. First, a numerical analysis of the approximation error resulting from the application of PFEM on the beam equations is carried out. This is followed by modeling the structural dynamics of a high-rise building composed of beam elements and an example of a multi-domain system with a hydraulic actuator. Concluding remarks are given in Sec.\,\ref{sec:conclusion}.
The models and methods of Secs.\,\ref{sec:truss_elements}\,-\,\ref{sec:assembly} were compiled into a Matlab framework which can be used for the modeling of arbitrary truss structures and frames as port-Hamiltonian systems. It is available for download on GitHub\footnote{\url{http://github.com/awarsewa/ph_fem/}}. 

\section{Beam Element Models}\label{sec:truss_elements}
Models of complex civil engineering structures are usually built by interconnecting basic element types with different parameters. Usually, the number of elements forming a model is much higher than the number of basic types it is composed of. In this article, we restrict ourselves to truss structures composed of rods and frames composed of beam elements. Shell elements are not yet included in the framework presented in this work. 
	\begin{figure}
		\centering
		\def\svgwidth{0.7\textwidth}
		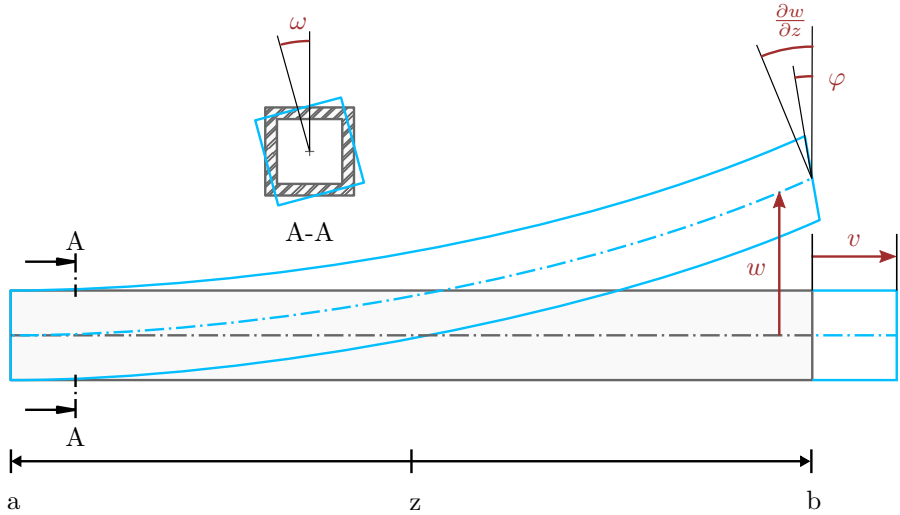
		\caption[Beam element]{A beam element defined on the interval $Z = \begin{bmatrix} a & b \end{bmatrix}$ and its deflections (blue) due to bending ($w$, $\varphi$), torsional ($\omega$) and axial stress ($v$). }%
		\label{fig:beam}%
	\end{figure}
Fig.\,\ref{fig:beam} depicts a beam defined on the spatial domain $Z := \begin{bmatrix} a & b \end{bmatrix}$ with spatial coordinate $z$ which is subject to bending, torsional and axial stress. When subject to bending, the beam experiences a deflection $w(z)$ from the neutral axis and a rotation of the cross section by the angle $\varphi(z)$. For torsional loads, the cross section rotates by the angle $\omega(z)$ about the neutral axis. Axial strain is determined by the deflection under axial loads, $v(z)$. Since the load cases are considered linearly independent, their dynamics can be treated separately. A complete beam model thus consists of three basic components which are presented in the following. For bending, a port-Hamiltonian model is derived for both the Euler-Bernoulli and the Timoshenko beam equations. The latter are used when the classic assumption that the cross-section is always perpendicular to the neutral axis, i.\,e. $\varphi = \partial w/\partial z$, does not hold (e.\,g. for thick beams). In Sec.\,\ref{subsec:link}, the model for a rod element deforming due to axial loads is presented. Torsion is modeled according to the Saint-Venant equations in Sec.\,\ref{subsec:saint-venant}. 

All component models are given as distributed parameter port-Hamiltonian systems in this section. Concentrated parameter models are obtained from the infinite dimensional systems via the procedure explained in Sec.\,\ref{sec:discretization}.

\subsection{Euler-Bernoulli Beam}\label{subsec:bernoulli}
In classical mechanics, the Euler-Bernoulli equation is widely used for describing the dynamics of a beam under bending stress. Cordoso-Ribeiro et al. \cite{cardoso2016piezoelectric} derived a port-Hamiltonian model for a beam actuated by piezoelectric patches using Euler-Bernoulli theory. We use the same formulation for the distributed parameter system neglecting the actuator terms. The Euler-Bernoulli equation is given as
	\begin{equation}\label{eq:bernoulli}
		\mu \frac{\partial^2 w}{\partial t^2} = - \frac{\partial^2}{\partial z^2} \left( E I \frac{\partial^2 w}{\partial z^2} \right),
	\end{equation}
	with the mass per unit length $\mu$, Young's modulus $E$ and the second moment of area $I$. For the formulation of the system Hamiltonian, two conjugate energy variables are chosen as
	\begin{equation}
		p^\mathrm A := \mu \frac{\partial w}{\partial t} , \qquad q^\mathrm A := \frac{\partial^2 w}{\partial z^2},
	\end{equation} 
The letter A is used to distinguish between the variables in this system of equations and their respective counterparts in others. With $x^\mathrm A = \begin{bmatrix} p^\mathrm A & q^\mathrm A \end{bmatrix}$, the energy of system~A can be stated in terms of it's Hamiltonian
	\begin{align}\label{eq:H_A}
		H^\mathrm A(x^\mathrm A) &= \int_{Z} \mathcal H^\mathrm A(x^\mathrm A) = \frac{1}{2} \int_{a}^b \left[ \frac{1}{\mu} (p^\mathrm A)^2\right] \mathrm{d}z + \frac{1}{2} \int_{a}^b \left[ EI (q^\mathrm A)^2 \right] \mathrm{d}z.
	\end{align}
For all the mechanical systems modeled in this contribution, the Hamiltonian can be separated into a kinetic energy part and a potential energy part. Here, $p^\mathrm A$ accounts for the former and $q^\mathrm A$ for the latter form of energy. Taking the variational derivative of $H^\mathrm A$ with respect to the energy variables yields the effort variables
		\begin{align}
				e^\mathrm A_p := \frac{\delta H^\mathrm A}{\delta p^\mathrm A} = \frac{\partial w}{ \partial t}, \qquad e^\mathrm A_q := \frac{\delta H^\mathrm A}{\delta q^\mathrm A} = E I \frac{\partial^2 w}{\partial z^2}
		\end{align}
where $e^\mathrm A_p(z, t)$ is the deflection velocity and $e^\mathrm A_{q}(z, t)$ the bending moment distributed over the interval $Z$. Defining the flow variables as $f^\mathrm A(z, t) := \dot{x}^\mathrm A$, system~A can be written as a distributed parameter Hamiltonian system of two conservation laws
	\begin{equation}\label{eq:pH_A}
	 f^\mathrm A =	\begin{bmatrix} f_p^\mathrm A \\ f_q^\mathrm A \end{bmatrix} = \underbrace{\begin{bmatrix} 
 0 & -\partial_{z^2}^2 \\
 \partial_{z^2}^2 & 0
\end{bmatrix}}_{\mathcal J^\mathrm A} \begin{bmatrix} e_p^\mathrm A \\ e_q^\mathrm A \end{bmatrix}.
	\end{equation}
The operator $\mathcal J^\mathrm A$ is formally skew symmetric according to \cite{zwart2009distributed} and therefore energy conserving. Substituting the definitions of flow and effort variables into \eqref{eq:pH_A} yields the Euler-Bernoulli equation \eqref{eq:bernoulli} and an equality relation. The energy flow of this system is given by
\begin{equation}\label{eq:dHdt_A}
	\begin{split} \dot H^\mathrm A &= \int_Z \frac{\delta H^\mathrm A}{\delta x^\mathrm A} \frac{\partial x^\mathrm A}{\partial t} \,\mathrm d z = \int_Z e_p^\mathrm A \dot x_p^\mathrm A + e_q^\mathrm A \dot x_q^\mathrm A \,\mathrm d z = \int_Z e_q^\mathrm A \partial^2_{z^2}e_p^\mathrm A- e_p^\mathrm A \partial^2_{z^2} e_q^\mathrm A \,\mathrm d z \\ 
	&= \int_Z \partial_z e_p^\mathrm A \partial_z e_q^\mathrm A - \partial_z e_q^\mathrm A \partial_z e_p^\mathrm A \,\mathrm d z + \Big[ e_q^\mathrm A \partial_z e_p^\mathrm A - e_p^\mathrm A \partial_z e_q^\mathrm A \Big]_a^b.
	\end{split}
\end{equation}
From the second line of \eqref{eq:dHdt_A}, we see that the energy flow depends only on the boundary values of the effort variables and their derivatives. This motivates the definition of boundary inputs $\vec u_\partial^\text A$ and outputs $\vec y_\partial^\text A$ such that the Hamiltonian rate of change can be expressed as $\dot H^\mathrm A = (\vec u_\partial^\text A)^\text T \vec y_\partial^\text A$. A possible choice of the boundary variables is given by
\begin{equation}\label{eq:boundary_A}
	\vec u_\partial^\text A = \begin{bmatrix} \partial_z e_q^\text A(a) & -\partial_z e_q^\mathrm A(b) & -e_q^\mathrm A(a) & e_q^\mathrm A(b) \end{bmatrix}^\text T, \quad \vec y_\partial^\text A = \begin{bmatrix} e_p^\mathrm A(a) & e_p^\mathrm A(b) & \partial_z e_p^\mathrm A(a) & \partial_z e_p^\mathrm A(b) \end{bmatrix}^\text T,
\end{equation}
where the boundary inputs are forces and moments and the boundary outputs the collocated velocities and angular velocities. As stated in e.\,g. \cite{le2005dirac}, other definitions of the boundary variables are also valid. However, as we will show in Sec.\,\ref{sec:discretization}, \eqref{eq:boundary_A} is a ``natural'' choice as it has a normalizing property in the application of PFEM. The definition of $\vec u_\partial^\text A$ and $\vec y_\partial^\text A$ completes the formulation of \eqref{eq:bernoulli} as distributed parameter port-Hamiltonian system with power exchange over the boundary.
	
	\subsection{Timoshenko Beam}\label{subsec:timoshenko}
	For thick beams, Eq.\,\eqref{eq:bernoulli} does not accurately describe the deformation under bending. The rotation of the cross section with respect to the mid-surface normal needs to be taken into account. Including the rotational dynamics, bending dynamics are described with a pair of equations in Timoshenko beam theory 
	\begin{equation}
		\begin{split}
			\rho A \frac{\partial^2 w}{\partial t^2} &= \frac{\partial}{\partial z} \left[ \kappa A G_\mathrm s \left(\frac{\partial w}{\partial z} - \varphi \right) \right] \\
	\rho I \frac{\partial^2 \varphi}{\partial t^2} &= \frac{\partial}{\partial z} \left( E I \frac{\partial \varphi}{\partial z} \right) + \kappa A G_\mathrm s \left(\frac{\partial w}{\partial z} - \varphi \right).
		\end{split}
		\end{equation}
			As in \eqref{eq:bernoulli}, $E$ and $I$ are Young's modulus and the second moment of area, respectively. The beam's cross sectional area is denoted by $A$, $G_\mathrm s$ is the shear modulus and $\kappa$ is the Timoshenko shear coefficient. A port-Hamiltonian formulation of the equations above can be found e.\,g. in \cite{macchelli2004modeling}. We proceed accordingly by defining a set of energy variables
		\begin{align}
				p^{\mathrm B}_1 := \rho I \frac{\partial \varphi}{\partial t}, \quad p^{\mathrm B}_2 := \rho A \frac{\partial w}{\partial t}, \qquad q^{\mathrm B}_1 := \frac{\partial \varphi}{\partial z}, \quad q^{\mathrm B}_2 := \left( \frac{\partial w}{\partial z} - \varphi \right)
		\end{align}
		which are collected in the state vector $x^\mathrm B := \begin{bmatrix} p_1^\mathrm B & p_2^\mathrm B & q_1^\mathrm B & q_2^\mathrm B \end{bmatrix}$. With the previous definitions, the system Hamiltonian can be written as
		\begin{align}
			H^\mathrm B(x^\mathrm B) &= \frac{1}{2} \int_{a}^b \left[\frac{1}{\rho I} (p^{\mathrm B}_1)^2 + \frac{1}{\rho A} (p^{\mathrm B}_2)^2 \right] \mathrm{d}z + \frac{1}{2} \int_{a}^b \left[ E I (q^{\mathrm B}_1)^2 + \kappa A G (q^{\mathrm B}_2)^2 \right] \mathrm{d}z.
		\end{align}
		Again, $p_1^\mathrm B$ and $p_2^\mathrm B$ are the kinetic energy variables of system B whereas $q_1^\mathrm B$ and $q_2^\mathrm B$ appear in potential energy terms. Parallel to the procedure in Sec.\,\ref{subsec:bernoulli}, we take the variational derivative of $H^\mathrm B$ with respect to $x^\mathrm B$ to obtain the effort variables for the Timoshenko beam
		\begin{align}
			\begin{split} 
				e^\mathrm B_{p1} &:= \frac{\partial \varphi}{\partial t}, \qquad e^\mathrm B_{q1} := E I \frac{\partial \varphi}{\partial z} \\
				e^\mathrm B_{p2} &:= \frac{\partial w}{\partial t}, \qquad e^\mathrm B_{q2} := \kappa A G_\mathrm s \left(\frac{\partial w}{\partial z} - \varphi \right).
			\end{split}
		\end{align}
Here, $e^\mathrm B_{p1}$ denotes the distributed angular velocity and $e^\mathrm B_{p2}$ the deflection velocity whereas $e^\mathrm B_{q1}$ is the distributed bending moment and $e^\mathrm B_{q2}$ the shear force. An infinite-dimensional Hamiltonian system of two conservation laws is obtained by defining the flow variables as $f^\mathrm B := \dot x^\mathrm B$ and establishing their relation to the effort variables as
		\begin{equation}\label{eq:pH_B}
 			f^\mathrm B = \begin{bmatrix} f_{p1}^\mathrm B \\ f_{p2}^\mathrm B \\ f_{q1}^\mathrm B \\ f_{q2}^\mathrm B \end{bmatrix} = \begin{bmatrix} 0 & 0 & \partial_z & 1 \\ 0 & 0 & 0 & \partial_z \\ \partial_z & 0 & 0 & 0 \\ -1 & \partial_z & 0 & 0 \end{bmatrix} \begin{bmatrix} e_{p1}^\mathrm B \\ e_{p2}^\mathrm B \\ e_{q1}^\mathrm B \\ e_{p2}^\mathrm B \end{bmatrix}.
		\end{equation}
	Compared to the Euler-Bernoulli beam, four equations instead of two are necessary to describe the dynamics of the Timoshenko beam in Hamiltonian form, which results in higher-order concentrated parameter systems. However, the second-order spatial derivative is not present in \eqref{eq:pH_B}. A convenient choice of the boundary variables for the Timoshenko beam is
\begin{equation}
		\vec u_\partial^\text B = \begin{bmatrix} -e_{q1}^\text B(a) & e_{q1}^\mathrm B(b) & -e_{q2}^\mathrm B(a) & e_{q2}^\mathrm B(b) \end{bmatrix}^\text T, \quad \vec y_\partial^\text B = \begin{bmatrix} e_{p1}^\mathrm B(a) & e_{p1}^\mathrm B(b) & e_{p2}^\mathrm B(a) & e_{p2}^\mathrm B(b) \end{bmatrix}^\text T.
\end{equation}
It is easy to verify that the Hamiltonian rate of change can be expressed as $\dot H^\mathrm B = (\vec u_\partial^\mathrm B)^\text T \vec y_\partial^\mathrm B$. 
	
	\subsection{Rod Element} \label{subsec:link}
	Modeling the dynamics of a linear beam is not complete without considering axial deformation. For a rod element which only supports axial loads, the deformation along the neutral axis $v(z)$ can be described by the following dynamic equation
		\begin{equation}\label{eq:link}
			\mu \frac{\partial^2 v}{\partial t^2} = \frac{\partial}{\partial z}\left(E A \frac{\partial v}{\partial_z} \right),
		\end{equation}
		with the mass per unit length $\mu$, Young's modulus $E$ and the beam's cross sectional area $A$ as defined in the preceding sections. In accordance with the procedure for Euler-Bernoulli and Timoshenko beam, conjugate energy variables are defined next
		\begin{equation}
			p^\mathrm C := \mu \frac{\partial v}{\partial t}, \quad q^\mathrm C := \frac{\partial v}{\partial z}.
		\end{equation}
		They separate the kinetic and potential energy domains and are used in the formulation of the system Hamiltonian
		\begin{align}
			H^\mathrm C(x^\mathrm C) &= \frac{1}{2} \int_{a}^b \left[ \frac{1}{\mu} (p^\mathrm C)^2\right] \mathrm{d}z + \frac{1}{2} \int_a^b \left[ E A (q^\mathrm C)^2 \right] \mathrm{d}z.
		\end{align}
		For the definition of the effort variables, we take the variational derivative of $H^\mathrm C$ with respect to $p^\mathrm C$ and $q^\mathrm C$
		
		\begin{align}
				e^\mathrm C_p := \frac{\partial v}{\partial t}, \qquad e^\mathrm C_q := E A \frac{\partial v}{\partial z},
		\end{align}
		where $e^\mathrm C_p$ is the axial deflection velocity and $e^\mathrm C_q$ the axial force distributed over the interval $Z$.
		By defining the flow variables as $f^\mathrm C := \dot x^\mathrm C$ we arrive at a formulation of rod element as infinite-dimensional Hamiltonian system of two conservation laws
		\begin{equation}\label{eq:pH_C}
			f^\mathrm C = \begin{bmatrix} f_p^\mathrm C \\ f_q^\mathrm C \end{bmatrix} = \begin{bmatrix} 
 0 & \partial_{z} \\
 \partial_{z} & 0
 \end{bmatrix} \begin{bmatrix} e_p^\mathrm C \\ e_q^\mathrm C \end{bmatrix}.
		\end{equation}
		Again, we obtain a representation as distributed parameter port-Hamiltonian system by selecting boundary inputs and outputs from the evaluation of the effort variables at the boundary
	\begin{equation}
		\vec u_\partial^\text C = \begin{bmatrix} -e_q^\text C(a) & e_q^\mathrm C(b) \end{bmatrix}^\text T, \quad \vec y_\partial^\text C = \begin{bmatrix} e_p^\mathrm C(a) & e_p^\mathrm C(b) \end{bmatrix}^\text T,
		\end{equation}
		which allows us to express the energy flow over the boundary as $\dot H^\mathrm C = (\vec u_\partial^\mathrm C)^\text T \vec y_\partial^\mathrm C$.
		
	\subsection{Saint-Venant Torsion Element}\label{subsec:saint-venant}
		Beam deformation due to torsion about the neutral axis is modeled using the Saint-Venant equation. Warping torsion is neglected here, assuming that cross sections are either warping-free or that warping is not constrained. Saint-Venant torsion is described by 
		\begin{equation}\label{eq:torsion}
			I_\mathrm p \frac{\partial^2 \omega}{\partial t^2} = \frac{\partial}{\partial z} \left(G_\mathrm s J_\mathrm t \frac{\partial \omega}{\partial_z}\right),
		\end{equation}
		where $I_\mathrm p$ is the polar moment of inertia, $G_\mathrm s$ the shear modulus, $J_\mathrm t$ the torsion constant and $\omega(z)$ the torsional angle as depicted in Fig.\,\ref{fig:beam}. We note that \eqref{eq:torsion} and \eqref{eq:link} are, except for the choice of parameters, identical equations. Consequently, the port-Hamiltonian formulation will be identical to \eqref{eq:pH_C} when choosing $p^\mathrm D := I_\mathrm p \partial \omega/\partial t$ and $q^\mathrm D := \partial \omega/\partial z$ as the energy variables. The reader is referred to the previous section or to \cite{cardoso2015modeling} for the remaining steps to obtain \eqref{eq:pH_C}.

\section{Spatial Discretization}\label{sec:discretization}
For the simulation of the dynamic response of beam elements, numerical approximation of the infinite dimensional port-Hamiltonian system equations derived in Sec.\ref{sec:truss_elements} is necessary. When employing spatial discretization methods, care must be taken to preserve the port-Hamiltonian structure, which is why conventional approaches are not readily applicable. A spatial discretization method is structure preserving if the resulting concentrated parameter system can be written in ISO form \eqref{eq:ISO_form}. We use PFEM, as proposed by Cardoso-Ribeiro et al. \cite{cardoso2019partitioned}, for reasons outlined in the introduction. In particular, due to its straightforward applicability and compatibility with standard FEM methods and software. In the following, the application of PFEM to systems of the form
\begin{equation}\label{eq:inf_class}
	\dot x = f = \mathcal J e, \quad \mathrm{with}~ \mathcal J = \sum_{i=0}^N \begin{bmatrix} 0 & A_i \\ (-1)^{i+1}A_i^\mathrm{T} & 0 \end{bmatrix} \frac{\partial^i}{\partial z^i}
\end{equation}
is presented. The matrices $A_i$ are quadratic with entries being either zero or one and $N$ is the maximum order of the spatial derivative. 
The system Hamiltonian can be expressed as 
\begin{equation} \label{eq:hamiltonian}
H(x) = \frac{1}{2} \int_Z x^\mathrm{T} \mathcal L x \,\mathrm{d}z,
\end{equation}
where $\mathcal L$ can be a function of $x$, and the interval $Z = \begin{bmatrix} a & b\end{bmatrix}$ as above. When observing the infinite dimensional port-Hamiltonian systems in Sec.\,\ref{sec:truss_elements}, we notice that they are all of this type with the order of the spatial derivative $N \leq 2$ and $\mathcal L$ independent of $x$ for isotropic beams. Taking for instance \eqref{eq:pH_B}, the Timoshenko beam equations in port-Hamiltonian form with $N = 1$, we get
\begin{equation}
	A_0 = \begin{bmatrix} 0 & 1 \\ 0 & 0 \end{bmatrix}, \quad A_1 = \begin{bmatrix} 1 & 0 \\ 0 & 1 \end{bmatrix}.
\end{equation} 
By treating the spatial discretization process on a system class level, explicit formulation of the process for every single load case of the beam is avoided. At the same time, it is more evident that the method's applicability is not limited to truss structures and frames.
	\begin{figure}
		\centering
		\def\svgwidth{0.7\textwidth}
		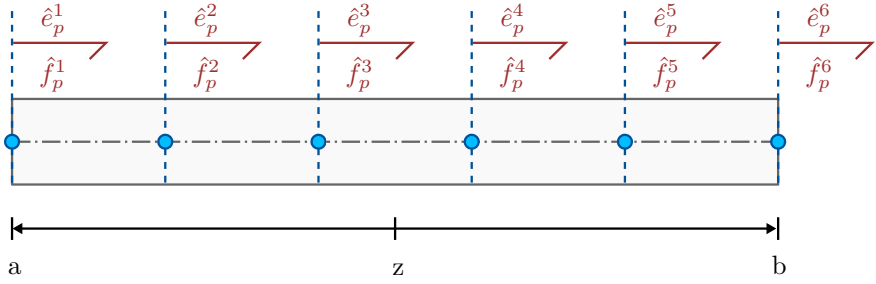
		\caption[Numerical approximation of a beam element]{Division of a beam element into five intervals with $N_p = 6$ supporting points with corresponding flows and efforts for numerical approximation. }%
		\label{fig:approximation}%
	\end{figure}

In a first step, the energy variables $x$, flow variables $f$ and effort variables $e$ of a system of class \eqref{eq:inf_class} are approximated using the method of weighted residuals
		\begin{align}\label{eq:approx}
			\begin{split}
				x_p(z, t) &\approx \hat{\vec x}_p^\mathrm{T}(t) \boldsymbol \phi_p(z), \quad x_q(z, t) \approx \hat{\vec x}_q^\mathrm{T}(t) \boldsymbol \phi_q(z), \\
				f_p(z, t) &\approx \hat{\vec f}_p^\mathrm{T}(t) \boldsymbol \phi_p(z), \quad f_q(z, t) \approx \hat{\vec f}_q^\mathrm{T}(t) \boldsymbol \phi_q(z), \\
				e_p(z, t) &\approx \hat{\vec e}_p^\mathrm{T}(t) \boldsymbol \phi_p(z), \quad e_q(z, t) \approx \hat{\vec e}_q^\mathrm{T}(t) \boldsymbol \phi_q(z).
			\end{split}
		\end{align}
		Here, the basis functions $\boldsymbol \phi_p(z)$ and $\boldsymbol \phi_q(z)$ are chosen as the Lagrange polynomials of order $N_p - 1$ and $N_q - 1$ respectively. The indices $p$ and $q$ are used to distinguish between kinetic and potential energy related terms. The vectors $\hat{\vec x}_p, \hat{\vec f}_p, \hat{\vec e}_p \in \mathbb R^{N_p}$ and $\hat{\vec x}_q, \hat{\vec f}_q, \hat{\vec e}_q \in \mathbb R^{N_q}$ are the energy variables, flows and efforts evaluated at uniformly distributed points in $Z = \begin{bmatrix} a & b\end{bmatrix}$. For values of $N_p \geq 2$ and $N_q \geq 2$, the boundary is always included. This is illustrated for the kinetic variables of a beam element in Fig.\,\ref{fig:approximation} where $N_p = 6$ supporting points are used. The approximations of kinetic efforts $\hat e_p^i$ and flows $\hat f_p^i$ are included in the figure with the half arrow pointing in the positive direction respectively. In order to be compatible with FEM-methods, PFEM proceeds via the weak form of \eqref{eq:inf_class}
		\begin{equation}\label{eq:weak_form}
			\int_Z \boldsymbol \phi(z) \vec f \;\mathrm{d}z = \int_Z \boldsymbol \phi(z) \mathcal J \vec e \;\mathrm{d}z 
		\end{equation}
using the approximation bases $\boldsymbol \phi_p(z)$ and $\boldsymbol \phi_q(z)$ as test functions. This transformation is similar to the Galerkin method, which is widely used for numerical approximation in FEM (see e.\,g. \cite{schnell2006technische, hughes2012finite}). Substituting the approximations \eqref{eq:approx} for energy variables, flows and efforts into \eqref{eq:weak_form} yields
		\begin{align}\label{eq:weak_system}
		\begin{split}
			\left( \int_Z \boldsymbol \phi_p \boldsymbol \phi_p^\mathrm{T} \mathrm{d}z \right) \hat{\vec f}_{p} &= \left( \sum_{i=0}^N \int_Z \boldsymbol \phi_p A_i \frac{\partial^i}{\partial z^i} \boldsymbol \phi_q^\mathrm{T} \mathrm{d}z \right) \hat{\vec e}_q \\
			\left( \int_Z \boldsymbol \phi_q \boldsymbol \phi_q^\mathrm{T} \mathrm{d}z \right) \hat{\vec f}_{q} &= 
			\left( \sum_{i=0}^N (-1)^{i+1} \int_Z \boldsymbol \phi_q A_i^\mathrm{T} \frac{\partial^i}{\partial z^i} \boldsymbol \phi_p^\mathrm{T} \mathrm{d}z \right) \hat{\vec e}_p.
		\end{split}
		\end{align}
For systems of two conservation laws such as \eqref{eq:inf_class}, structure-preserving mixed Galerkin approximation as presented in \cite{kotyczka2018weak} can be used. However, with PFEM, the finite-dimensional port-Hamiltonian system can be obtained in a direct fashion without further projection. This is achieved by applying integration by parts to only one of the equations in \eqref{eq:weak_system}. In principle, either equation can be integrated by parts. By choosing one or the other, we decide on the causality of the boundary ports -- i.\,e. which variables will be the boundary inputs and which the boundary outputs. When integrating the respective dual equation by parts instead, causality of the boundary ports is reversed. For the systems defined in Sec.\,\ref{sec:truss_elements}, integrating the upper equation by parts results in forces and torques as boundary inputs and velocities and angular velocities as boundary outputs. Integrating the lower equation by parts instead, the role of boundary inputs and outputs is reversed. To be consistent with the FEM approach, we choose to apply integration by parts N times to the equation containing the kinetic flows. Intermediate steps of this calculation are shown in \ref{app:approx}. As a result we obtain 
	\begin{align}\label{eq:partial}
		\begin{split}	\left( \int_Z \boldsymbol \phi_p \boldsymbol \phi_p^\mathrm{T} \mathrm{d}z \right) \hat{\vec f}_{p} = & \left( \sum_{i=0}^N (-1)^i \int_Z \frac{\partial^i}{\partial z^i}\boldsymbol \phi_p A_i \boldsymbol \phi_q^\mathrm{T} \mathrm{d}z \right) \hat{\vec e}_q \\ & + \sum_{i=1}^{N} \sum_{j=1}^i (-1)^{j-1} \left[ \frac{\partial^{j-1}}{\partial z^{j-1}} \boldsymbol \phi_p A_i \frac{\partial^{i-j}}{\partial z^{i-j}}\boldsymbol \phi_q^\mathrm{T} \hat{\vec e}_q \right]_a^b.
		\end{split}
	\end{align}
Observing the term in round brackets on the right hand side, we note that it is the negative transpose of the term in front of $\hat{\vec e}_p$ in the lower equation in \eqref{eq:weak_system}. Applying Gaussian quadrature to evaluate the integrals, a concentrated parameter approximation of the system dynamics is obtained
	\begin{equation} \label{eq:sys_gauss}
		\underbrace{\begin{bmatrix} \vec M_p & 0 \\ 0 & \vec M_q \end{bmatrix}}_{\vec M} \hat{\vec f} = \underbrace{\begin{bmatrix} 0 & \vec P \\ -\vec P^\mathrm{T} & 0 \end{bmatrix}}_{\vec J} \hat{\vec e} + \underbrace{\begin{bmatrix} \vec G_p \\ 0 \end{bmatrix}}_\vec G \vec u_\partial
	\end{equation}
	with the symmetric mass matrices $\vec M_p \in \mathbb R^{N_p\times N_p}$ and $\vec M_q \in \mathbb R^{N_q \times N_q}$, the square skew-symmetric interconnection matrix $\vec J$ with $\vec P \in \mathbb R^{N_p \times N_q}$, the input matrix $\vec G$ and the boundary input variables $\vec u_\partial$. The latter are the efforts $\boldsymbol \phi_q^\mathrm{T} \vec e_q$ and their spatial derivatives, appearing in the second term on the right hand side in \eqref{eq:partial}, evaluated at the boundary. According to the effort definitions in Sec.\,\ref{sec:truss_elements} in case of the beam dynamics, the inputs are forces and bending moments. For one-dimensional systems such as the beam elements considered in this publication, the number of boundary inputs amounts to $2 N$ with $\vec u_\partial \in \mathbb R^{2 N}$ and $\vec G_p \in \mathbb R^{N_p \times 2N}$. With the boundary inputs stated for the individual systems in Sec.\,\ref{sec:truss_elements}, it can be easily verified that $\vec G_p$ is directly obtained by evaluation of $\boldsymbol \phi_p$ (and $\partial_z \boldsymbol \phi_p$ in case of the Euler-Bernoulli beam) at the boundary. This is why we regard them as ``natural'' choices for the application of PFEM. For the systems B, C and D, we thus get $\vec G_p = \vec I$ for $N_p = 2N$.  

Equation \eqref{eq:sys_gauss} is already a valid finite-dimensional approximation of the port-Hamiltonian system, but it is not yet in the desired input-state-output form \eqref{eq:ISO_form}. To bring it to this form, we recall that $f := \dot x$ and transform the energy variables
	\begin{equation} \label{eq:state_transform}
		\tilde{\vec x} := \vec M \hat{\vec x}.
	\end{equation}
For a concentrated-parameter port-Hamiltonian system, the effort variables are obtained by taking the gradient of the system Hamiltonian with respect to the energy variables. Thus, it remains to show that
\begin{equation} \label{eq:assumption}
	\hat{\vec e} = \nabla_{\tilde{\vec x}} H(\tilde{\vec x}).
\end{equation} 
First, the system Hamiltonian needs to be written in terms of $\tilde{\vec x}$ which is achieved by substitution of the approximations \eqref{eq:approx} into \eqref{eq:hamiltonian} and application of the transformation \eqref{eq:state_transform} 
\begin{equation}
H(\tilde{\vec x}) = \frac{1}{2} \tilde{\vec x}^\mathrm{T} \vec M^{-1} \left(\int_Z \boldsymbol \phi \mathcal L \boldsymbol \phi^\mathrm{T} \mathrm{d}z \right) \vec M^{-1} \tilde{\vec x}.
\end{equation}
From Sec.\,\ref{sec:truss_elements}, we remember that $e = \partial H/\partial x$. After substitution of the approximations \eqref{eq:approx} and transformation to the weak form, this expression becomes
\begin{equation}
	\left(\int_Z \boldsymbol \phi \boldsymbol \phi^\mathrm{T} \mathrm{d}z \right) \hat{\vec e} = \left(\int_Z \boldsymbol \phi \mathcal L \boldsymbol \phi^\mathrm{T} \mathrm{d}z \right) \hat{\vec x}.
\end{equation}
Since the integral on the left hand side is $\vec M$, it is easily seen that \eqref{eq:assumption} is true. We define
\begin{equation}
 \vec Q := \vec M^{-1} \int_Z \boldsymbol \phi \mathcal L \boldsymbol \phi^\mathrm{T} \mathrm{d}z 
\end{equation}
such that system \eqref{eq:sys_gauss} can be written as
	\begin{align}\label{eq:discrete_system}
	\begin{split}
		\dot{\tilde{\vec x}} &= \vec J \vec Q \tilde{\vec x} + \vec G \vec u_\partial \\
		\vec y_\partial &= \vec G^\mathrm{T} \vec Q \tilde{\vec x}
		\end{split}
	\end{align}
with the boundary output variables $\vec y_\partial$ being the input-collocated velocities and angular velocities for all systems considered in this article.

\section{Assembly of Complex Systems}\label{sec:assembly}
	\begin{figure}
		\centering
		\def\svgwidth{0.75\textwidth}
		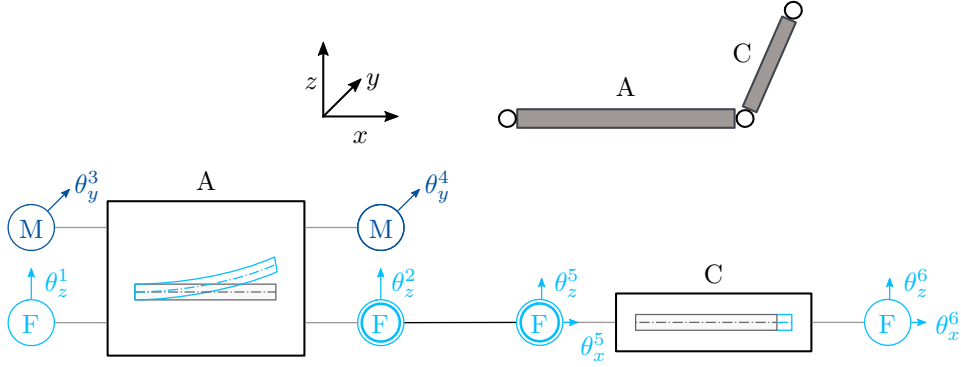
		\caption{Interconnection of two linear elastic elements with boundary mechanical ports.}%
		\label{fig:assembly}%
	\end{figure}
With the systems defined in Sec.\,\ref{sec:truss_elements} and the numerical approximation method described in Sec.\,\ref{sec:discretization}, the dynamics of one-dimensional primitive elements can be obtained in ODE form \eqref{eq:discrete_system} for all load cases of a linear beam. To model more complex truss structures and frames, the interconnection and coupling of multiple basic elements needs to be taken into account. In the following, a method for the automatic assembly of arbitrary 3D structures composed of beams and rods is derived. Throughout this section we make use of Fig.\,\ref{fig:assembly} to illustrate the concepts. It depicts the interconnection of a Euler-Bernoulli beam and a rod element respectively labeled system ``A'' and ``C''. The system equations of the Euler-Bernoulli beam are
\begin{align}\label{eq:sys_A}
	\begin{split}
		\tilde{\vec f}^\mathrm A &= \vec J^\mathrm A \hat{\vec e}^\mathrm A + \vec G^\mathrm A \vec u_\partial^\mathrm A, \\
		\vec y_\partial^\mathrm A &= (\vec G^\mathrm A)^\mathrm{T} \hat{\vec e}^\mathrm A,
	\end{split}
\end{align}
with $\vec u_\partial^A \in \mathbb R^4$ the shear forces and bending moments at the boundary and $\vec y_\partial^A$ the corresponding collocated velocities and angular velocities. The dimension of $\tilde{\vec f}^A$ depends on the order of the approximation polynomials $\boldsymbol \phi_p$ and $\boldsymbol \phi_q$ in \eqref{eq:approx}. For the rod element we obtain
\begin{align}\label{eq:sys_C}
	\begin{split}
		\tilde{\vec f}^\mathrm C &= \vec J^\mathrm C \hat{\vec e}^\mathrm C + \vec G^\mathrm C \vec u_\partial^\mathrm C, \\
		\vec y_\partial^\mathrm C &= (\vec G^\mathrm C)^\mathrm{T} \hat{\vec e}^\mathrm C,
	\end{split}
\end{align}
where $\vec u_\partial^\mathrm C \in \mathbb R^2$ are the axial forces at the boundary and $\vec y_\partial^\mathrm C$ the collocated velocities.

In the bottom part of Fig.\,\ref{fig:assembly} the interconnection of A and C is depicted in a mechanical network diagram. Specific to this block diagram representation is the appearance of a normalized orientation vector $\boldsymbol \theta^i$ at each boundary port, where $i$ denotes the port number. A port labeled ``M'' indicates a torque/angular velocity pair at the boundary and ``F'' marks ports with a force/velocity pair. Ports of the same type with a non-zero scalar product of the orientation vectors can be coupled. Given the information in Fig.\,\ref{fig:assembly}, it is possible to write an algorithm for the automatic assembly of the coupled system. In the remainder of this section, the necessary steps such an algorithm needs to perform are explained in more detail. First, a way to concatenate the system matrices of individual elements is shown, followed by the generation of coupling constraints. Finally, the elimination of the generated constraints is outlined in Sec.\,\ref{subsec:elimination}.

	\subsection{System Concatenation}
	Without considering the coupling of the dynamic equations yet, there are several ways to concatenate the system matrices of systems A and C. When performing integration by parts on the equations involving the kinetic flows, the boundary inputs always act on $f_p$ and never on $f_q$. Thus, we propose to maintain the separation of kinetic and potential energy on concatenation. As we will later see, this also facilitates notation in further steps. For both \eqref{eq:sys_A} and \eqref{eq:sys_C}, the interconnection matrix and input matrix have the following structure
		\begin{equation}
		 \vec J = \begin{bmatrix} 0 & \vec J_p \\ \vec J_q & 0 \end{bmatrix}, \quad \vec G = \begin{bmatrix} \vec G_p \\ 0 \end{bmatrix}.
		\end{equation}
When defining the concatenated energy variables as $\tilde{\vec x}^\mathrm{AC} := \begin{bmatrix} \tilde{\vec x}_p^\mathrm C & \tilde{\vec x}_p^\mathrm A & \tilde{\vec x}_q^\mathrm A & \tilde{\vec x}_q^\mathrm C \end{bmatrix}^\mathrm{T}$, the resulting joint system matrices are 
		\begin{equation}\label{eq:sep}
		 \vec J^\mathrm{AC} = \begin{bmatrix} 0 & 0 & 0 & \vec J^{\mathrm C}_p \\ 0 & 0 & \vec J^{\mathrm A}_p & 0 \\ 0 & \vec J^{\mathrm A}_q & 0 & 0 \\ \vec J^{\mathrm C}_q & 0 & 0 & 0 \end{bmatrix}, \quad \vec G^\mathrm{AC} = \begin{bmatrix} 0 & \vec G^{\mathrm C}_p \\ \vec G^{\mathrm A}_p & 0 \\ 0 & 0 \\ 0 & 0 \end{bmatrix}
		\end{equation}
and the structural separation of energy domains is maintained.

	\subsection{Formulation of Constraint Equations}\label{subsec:const}
	According to Newton's second law of motion, forces and torques at each nodal point of a structure have to sum up to zero respectively for each degree of freedom considered. At the same time, velocities and angular velocities with the same orientation have to be identical. Looking again at the example in Fig.\,\ref{fig:assembly}, this implies 
	\begin{equation}\label{eq:sample_vel}
		v_b^\mathrm A = v_z^2, \quad v_a^\mathrm C = \theta_z^5 v_z^2
	\end{equation} 
	where $v_b^\mathrm A$ is the right boundary velocity in $z$-direction of system A, $v_a^\mathrm C$ the left boundary velocity in $z$-direction of the rod and $v_z^2$ the velocity in $z$-direction of the second node. The components of the orientation vector of the fifth port $\boldsymbol \theta^5$ are depicted in Fig.\,\ref{fig:assembly}. Since the Euler-Bernoulli beam cannot take loads in the $x$-direction, $\theta^5_x$ does not appear in the formulation of constraints. For the boundary forces $F^\mathrm A_b$ and $F_a^\mathrm C$ we get
	\begin{equation}\label{eq:sample_force}
			F_b^\mathrm A + \theta^5_z F_a^\mathrm C + F_z^2 = 0
	\end{equation}
	with the additional contribution of an external force $F_z^2$ in $z$-direction. 
	
	Since the boundary outputs of the concatenated system include all boundary velocities and angular velocities, we can write \eqref{eq:sample_vel} in a more general fashion. For an arbitrary structure, we define its vector of global nodal velocities $\dot{\vec q} \in \mathbb R^{n_\mathrm{DOF}}$ with $n_\mathrm{DOF}$ the total number of DOFs. It contains a velocity or angular velocity variable for each DOF. This way, we can express each boundary output as a linear combination of entries of $\dot{\vec q}$
		\begin{equation}\label{eq:vel_con}
			y^i_\partial = (\vec c^i_v)^\mathrm{T} \dot{\vec q}, \quad i = 1 \dots n_u
		\end{equation}
where $n_u$ is the total number of boundary inputs or outputs and $\vec c^i_v$ maps $\boldsymbol \theta^i$ to the corresponding nodal velocity or angular velocity. From the relation between $\dot{\vec q}$ and $\vec y_\partial$, algebraic constraints on the state variables $\vec x$ are obtained by eliminating $\dot{\vec q}$ 
		\begin{equation}
			\vec y_\partial = \vec C_v \dot{\vec q} \quad \Rightarrow \quad \vec C_v^\perp \vec y_\partial = 0
		\end{equation}
using a left annihilator $\vec C_v^\perp$ of $\vec C_v$ and the output equation of \eqref{eq:ISO_form}. Resulting algebraic constraints are of the form
		\begin{equation}\label{eq:output_constraints}
			0 = \vec B^\mathrm{T} \vec Q \tilde{\vec x}.
		\end{equation}
For statically determinate structures, the row rank of $\vec C_v^\perp$ is $n_\mathrm{DOF}$ and it follows that $\vec B \in \mathbb R^{n \times n_\mathrm c}$ with $n_\mathrm c = n_u - n_\mathrm{DOF}$. 

In a similar fashion, we can find a generalized expression for \eqref{eq:sample_force} and use it to reformulate the boundary inputs $\vec u_\partial$. For each global DOF, an external force or torque is introduced and they are collected in the vector of external forces and torques $\vec f_\mathrm{ext} \in \mathbb R^{n_\mathrm{DOF}}$. Consequently, each external force or torque can be equated with a linear combination of boundary inputs
		\begin{align}
			\begin{split}
			(c^j_u)^\mathrm{T} \vec u_\partial &= f^j_\mathrm{ext}, \quad j = 1 \dots n_\mathrm{DOF} \\
			\vec C_u \vec u_\partial &= \vec f_\mathrm{ext}.
			\end{split}
		\end{align}
Usually, the number of DOFs is lower than the number of boundary inputs. With $n_\mathrm{DOF}$ input constraints, this means that free boundary inputs $\vec u_\mathrm f \in \mathbb R^{n_\mathrm c}$ can be chosen. This enables a reformulation of the input term $\vec G \vec u_\partial$ as
		\begin{equation}\label{eq:free_ext}
			\vec G \vec u_\partial = \vec G_\mathrm f \vec u_\mathrm{f} + \vec K \vec f_\mathrm{ext},
		\end{equation}
where $\vec G_\mathrm f$ is the resulting input matrix of the free inputs $\vec u_\text f$ and $\vec K$ that of the external forces $\vec f_ \text{ext}$. Assuming $\vec G_\mathrm f$ has full column rank, we can construct its left inverse $\vec G_\mathrm f^+ = (\vec G^\mathrm{T}_\mathrm f \vec G_\mathrm f)^{-1} \vec G^\mathrm{T}_\mathrm f$ and rewrite $\vec u_\mathrm f$ as
				\begin{equation}\label{eq:free_lambda}
				\vec u_\mathrm f = \vec G_\mathrm f^+ \vec B \boldsymbol \lambda 
			\end{equation}
where $\boldsymbol \lambda \in \mathbb R^{n_\mathrm c}$ are the Lagrange multipliers corresponding to the algebraic constraints \eqref{eq:output_constraints}. Given \eqref{eq:output_constraints}, \eqref{eq:free_ext} and \eqref{eq:free_lambda}, the coupled structure can be written as a DAE system
			\begin{align} \label{eq:DAE}
			\begin{split}
				\dot{\tilde{\vec x}} &= \vec J \vec Q \tilde{\vec x} + \vec K \vec f_\mathrm{ext} + \vec B \boldsymbol \lambda \\
				\dot{\vec q} &= \vec K^\mathrm{T} \vec Q \tilde{\vec x} \\
				0 &= \vec B^\mathrm{T} \vec Q \tilde{\vec x}
			\end{split}
			\end{align}
where the collocated outputs to the external force inputs are the global nodal velocities $\dot{\vec q}$ . It can be shown that the DAE system~\eqref{eq:DAE} is of index one in case the matrix $\vec B^\mathrm{T} \frac{\partial^2 H}{\partial \tilde{\vec x}^2} \vec B$ has full rank \cite{van2014port}, which is always the case for the systems considered here. For now, we do not make further use of the DAE formulation. Instead, in the next section, we show how to eliminate the algebraic constraints to obtain an ODE formulation of the structural dynamics. For a thorough treatise of differential algebraic port-Hamiltonian systems, see \cite{van2013port}.

	\subsection{Algebraic Constraint Elimination}\label{subsec:elimination}
	Systems of the form \eqref{eq:DAE} can be reduced to the constrained state space $\mathcal X_\mathrm c$ with uniquely defined dynamics. As a consequence, the algebraic constraints are eliminated. A procedure for doing so is described in detail in e.\,g. \cite{ribeiro2016phd} and \cite{wu2014port}. We repeat the necessary steps here, to give a complete description of the methodology used to obtain the dynamic equations for arbitrary structure models in ODE form.
	
In a first step, the transformation
		\begin{equation}\label{eq:trans}
			\vec V = \begin{bmatrix} \vec B^\perp \\ \vec B^+\end{bmatrix}
		\end{equation}
is introduced. It is composed of both the left annihilator $\vec B^\perp$ and the left inverse $\vec B^+$ of $\vec B$. Left-multiplying the uppermost equation in \eqref{eq:DAE} with $\vec V$ and defining a new state vector $\vec z := \vec V \tilde{\vec x}$, yields
		\begin{equation}
			 \dot{\vec z} = \vec V \vec J \vec Q \tilde{\vec x} + \vec V \vec K \vec f_\mathrm{ext} + \begin{bmatrix} 0 \\ \vec I \end{bmatrix} \boldsymbol \lambda
		\end{equation}
To retain the port-Hamiltonian form, the term $\vec Q \tilde{\vec x}$ needs to be replaced with the gradient of $H$ with respect to $\vec z$
		\begin{align}
			\begin{split}
			\nabla_{\vec z} H &= \left(\frac{\partial \tilde{\vec x}}{\partial \vec z}\right)^\mathrm{T} \nabla_{\tilde{\vec x}} H \\
												&= \vec V^{-\mathrm T} \vec Q \vec V^{-1} \vec z \\
												&= \tilde{\vec Q} \vec z 
			\end{split}
		\end{align}
Defining $\tilde{\vec J} := \vec V \vec J \vec V^\mathrm{T}$ and $\tilde{\vec K} := \vec V \vec K$ and separating $\vec z$ into $\vec z_1 \in \mathbb R^{n-n_\mathrm c}$ and $\vec z_2 \in \mathbb R^{n_\mathrm c}$, the system can be reformulated as follows
		\begin{align}
			\begin{split}
				\begin{bmatrix} \dot{\vec z}_1 \\ \dot{\vec z}_2 \end{bmatrix} &= \begin{bmatrix} \tilde{\vec J}_{11} & \tilde {\vec J}_{12} \\ \tilde{\vec J}_{21} & \tilde{\vec J}_{22} \end{bmatrix}\begin{bmatrix} \tilde{\vec Q}_{11} & \tilde{\vec Q}_{12} \\ \tilde{\vec Q}_{21} & \tilde{\vec Q}_{22} \end{bmatrix} \begin{bmatrix} \vec z_1 \\ \vec z_2 \end{bmatrix} + \begin{bmatrix} \tilde{\vec K}_1 \\ \tilde{\vec K}_2 \end{bmatrix} \vec f_\mathrm{ext} + \begin{bmatrix} 0 \\ \vec I \end{bmatrix} \boldsymbol \lambda \\
				\dot{\vec q} &= \begin{bmatrix} \tilde{\vec K}_1^\mathrm T & \tilde{\vec K}_2^\mathrm T \end{bmatrix} \begin{bmatrix} \tilde{\vec Q}_{11} & \tilde{\vec Q}_{12} \\ \tilde{\vec Q}_{21} & \tilde{\vec Q}_{22} \end{bmatrix} \begin{bmatrix} \vec z_1 \\ \vec z_2 \end{bmatrix}\\
				0 &= \begin{bmatrix} 0 & \vec I \end{bmatrix} \begin{bmatrix} \tilde{\vec Q}_{11} & \tilde{\vec Q}_{12} \\ \tilde{\vec Q}_{21} & \tilde{\vec Q}_{22} \end{bmatrix} \begin{bmatrix} \vec z_1 \\ \vec z_2 \end{bmatrix}.
			\end{split}
		\end{align}
The third equation implies $\nabla_{\vec z_2} H = 0$ and that $\vec z_2$ can be expressed as
		\begin{equation}
			\vec z_2 = -\tilde{\vec Q}^{-1}_{22} \tilde{\vec Q}_{21} \vec z_1.
		\end{equation}
Therefore, the system dynamics on the constrained state space $\mathcal X_\mathrm c$ are given as
		\begin{align}
			\begin{split}
				\dot{\vec z}_1 &= \tilde{\vec J}_{11} (\tilde{\vec Q}_{11} - \tilde{\vec Q}_{12}\tilde{\vec Q}^{-1}_{22} \tilde{\vec Q}_{21}) \vec z_1 + \tilde{\vec K}_{11} \vec f_\mathrm{ext}\\
				\dot{\vec q} &= \tilde{\vec K}_{1}^\mathrm{T} (\tilde{\vec Q}_{11} - \tilde{\vec Q}_{12}\tilde{\vec Q}^{-1}_{22} \tilde{\vec Q}_{21}) \vec z_1.
			\end{split}
		\end{align}
Due to the separation of energy and co-energy related terms according to \eqref{eq:sep}, the interconnection matrix $\tilde{\vec J}_{11}$ of the ODE system has the following structure
	\begin{equation}
		\tilde{\vec J}_{11} = \begin{bmatrix} 0 & \tilde{\vec J}_p \\ -\tilde{\vec J}_p^\mathrm{T} & 0 \end{bmatrix}
	\end{equation}
with $\tilde{\vec J}_p \in \mathbb R^{(n_p - n_c) \times (n - n_c)} $, where $n_p$ is the number of potential energy variables in the DAE system \eqref{eq:DAE} and $(n_p - n_c) = n_\text{DOF}$. Since the external forces $\vec f_\mathrm{ext}$ and the constraint forces $\boldsymbol \lambda$ in \eqref{eq:DAE} act on $\tilde{\vec x}_p$ only, the transformation \eqref{eq:trans} retains $\tilde{\vec x}_q$ in the new state $\vec z_1$.

\subsection{Elimination of Linearly Dependent States}\label{subsec:linear_dependency}
With $\tilde{\vec x}_q$ retained in the state vector $\vec z_1$, the system dynamics of the reduced order ODE system can be formulated as follows using $\vec z_1 = \begin{bmatrix} \vec z_p & \tilde{\vec x}_q \end{bmatrix}^\text T$
\begin{align}\label{eq:sys_z1}
		\begin{split}
			\begin{bmatrix} \dot{\vec z}_p \\ \dot{\tilde{\vec x}}_q \end{bmatrix} &= \begin{bmatrix} 0 & \tilde{\vec J}_p \\ -\tilde{\vec J}_p^\mathrm T & 0 \end{bmatrix}\begin{bmatrix} \tilde{\vec Q}_p & 0 \\ 0 & \vec Q_q \end{bmatrix} \begin{bmatrix} \vec z_p \\ \tilde{\vec x}_q \end{bmatrix} + \begin{bmatrix} \tilde{\vec K}_p \\ 0 \end{bmatrix} \vec f_\mathrm{ext}, \\
			\dot{\vec q} &= \begin{bmatrix} \tilde{\vec K}_p^\mathrm T & 0 \end{bmatrix} \begin{bmatrix} \tilde{\vec Q}_p & 0 \\ 0 & \vec Q_q \end{bmatrix} \begin{bmatrix} \vec z_p \\ \tilde{\vec x}_q \end{bmatrix}.
		\end{split}
\end{align}

In this representation, it becomes clear that the $n_q$ state derivatives $\dot{\tilde{\vec x}}_q$ are linearly dependent since they are computed using a reduced number of potential energy variables $\vec z_p \in \mathbb R^{n_p-n_c}$. For ease of notation, we assume $n_p = n_q$ in the following, without loss of generality as long as $n_q > n_p - n_c$. We can then proceed to construct an orthonormal basis $\vec T_q$ of $-\tilde{\vec J}_p^\text T$ such that
\begin{equation}\label{eq:orth_transform}
	\tilde{\vec x}_q = \vec T_q \vec z_q.
\end{equation}
The effort variables $\tilde{\vec e}_q$ are then related to $\vec z_q$ as follows 
\begin{equation}
	\tilde{\vec e}_q = \nabla_{\tilde{\vec x}_q} H(\vec z_1) = \vec Q_q \tilde{\vec x}_q = \vec Q_q \vec T_q \vec z_q.
\end{equation}
This allows a formulation of algebraic constraints on $\tilde{\vec e}_q$ using $\vec B^* = \operatorname{ker} \left[ (\vec Q_q \vec T_q)^\mathrm T \right]$ such that
\begin{equation}
	(\vec B^*)^\text T \vec Q_q \tilde{\vec x}_q = 0.
\end{equation}
Given this relation, the process outlined in the previous section can be repeated, which results in the elimination of linearly dependent efforts or rather state variables. Afterward, the system \eqref{eq:sys_z1} is expressed in the new coordinates $\bar{\vec z} = \begin{bmatrix} \vec z_p & \vec z_q \end{bmatrix}^\text T$ as
\begin{align}\label{eq:sys_z_bar}
	\begin{split}
			\dot{\bar{\vec z}} &= \bar{\vec J} \bar{\vec Q} \bar{\vec z} + \bar{\vec K} \vec f_\text{ext}, \\
			\dot{\vec q} &= \bar{\vec K}^\text T \bar{\vec Q} \bar{\vec z}
	\end{split}
\end{align}
with $\bar{\vec z} \in \mathbb R^{2 n_\text{DOF}}$. Note, that the reduced order system \eqref{eq:sys_z_bar} is of the same order as the second order mechanical system in ISO port-Hamiltonian form \eqref{eq:ph_iso_mech}, given the global DOFs $\vec q$ are identical. In the next section it is shown, how to transform \eqref{eq:sys_z_bar} to obtain \eqref{eq:ph_iso_mech} without dissipation.

\subsection{Transformation to Global Coordinates}\label{subsec:to_global}
In the following, we assume that \eqref{eq:ph_iso_mech} -- without the damping term -- and \eqref{eq:sys_z_bar} describe the same system. In this case, \eqref{eq:ph_iso_mech} with $\vec R = 0$ can be obtained from \eqref{eq:sys_z_bar} by means of two consecutive transformations. With the first transformation
\begin{equation}
	\vec T_u = \begin{bmatrix} \bar{\vec K}^+ \\ \bar{\vec K}^\perp \end{bmatrix} 
\end{equation}
a normalization of \eqref{eq:sys_z_bar} with respect to the input $\vec f_\text{ext}$ is achieved 
\begin{equation}\label{eq:applying_Tu}
	\vec T_u \dot{\bar{\vec z}} = \vec T_u \bar{\vec J} \vec T_u^\text T \vec T_u^{-\text T} \bar{\vec Q} \vec T_u^{-1} \vec T_u \bar{\vec z} + \vec T_u \bar{\vec K} \vec f_\text{ext}.
\end{equation}
Since $\vec T_u \bar{\vec K} = \begin{bmatrix}\vec I & 0 \end{bmatrix}^\text T$, it follows that $\vec T_u \dot{\bar{\vec z}} = \begin{bmatrix} \vec M_\text{fe} \ddot{\vec q} & \dot{\tilde{\vec z}}_q \end{bmatrix}^\text T$, which results in
\begin{align}\label{eq:half_transformed}
	\begin{split}
			\begin{bmatrix} \vec M_\text{fe} \ddot{\vec q} \\ \dot{\tilde{\vec z}}_q \end{bmatrix} &= \underbrace{\begin{bmatrix} 0 & \bar{\vec P} \\ -\bar{\vec P}^\mathrm T & 0\end{bmatrix}}_{\vec T_u \bar{\vec J} \vec T_u^\text T} \underbrace{\begin{bmatrix} \vec M_\text{fe}^{-1} & 0 \\ 0 & \bar{\vec S} \end{bmatrix}}_{\vec T_u^{-\text T} \bar{\vec Q} \vec T_u^{-1}} \begin{bmatrix} \vec M_\text{fe} \dot{\vec q} \\ \tilde{\vec z}_q \end{bmatrix} + \begin{bmatrix} \vec I \\ 0 \end{bmatrix} \vec f_\mathrm{ext}, \\
			\dot{\vec q} &= \begin{bmatrix} \vec I & 0 \end{bmatrix} \begin{bmatrix} \vec M_\text{fe}^{-1} & 0 \\ 0 & \bar{\vec S} \end{bmatrix} \begin{bmatrix} \vec M_\text{fe} \dot{\vec q} \\ \tilde{\vec z}_q\end{bmatrix}.
	\end{split}
\end{align}
The final step to obtain \eqref{eq:ph_iso_mech} is the construction of a second transformation
\begin{equation}
	\vec T_p = \begin{bmatrix} \vec I & 0 \\ 0 & (-\bar{\vec P}^\mathrm T)^+ \end{bmatrix}, \quad \text{with~} \vec T_p \vec T_u \bar{\vec J} \vec T_u^\text T \vec T_p^\text T = \begin{bmatrix} 0 & -\vec I \\ \vec I & 0 \end{bmatrix}.
\end{equation}
When $\vec T_p$ is applied to \eqref{eq:half_transformed} in the same fashion as $\vec T_u$ in \eqref{eq:applying_Tu}, it follows that the transformed system must be identical to \eqref{eq:ph_iso_mech} with $\vec R = 0$. In case $\vec D_\text{fe}$ in \eqref{eq:mech_2nd_order} is a Rayleigh or Caughey damping term, $\vec R$ can now be easily constructed from $\vec M_\text{fe}$ and $\vec K_\text{fe}$. Other types of damping might involve adding dissipation terms on the PDE level, but this is not considered here.

\section{Numerical Examples}\label{sec:example}
In this section, we provide numerical examples that illustrate the use of the presented methods. First, the approximation error of PFEM for the beam equations in Sec.\,\ref{sec:truss_elements} is analyzed numerically. This is followed by a dynamic model of a high-rise building composed of many different elements which illustrates all of Sec.\,\ref{sec:assembly}. Finally, the coupling of a mechanical system with hydraulic cylinders (i.\,e. a multi-domain system) is presented.

\subsection{Analysis of Approximation Error}
\begin{figure}
\begin{subfigure}[l]{0.5\textwidth} 
	\centering
	\includegraphics[width=0.95\textwidth, height=6.5cm]{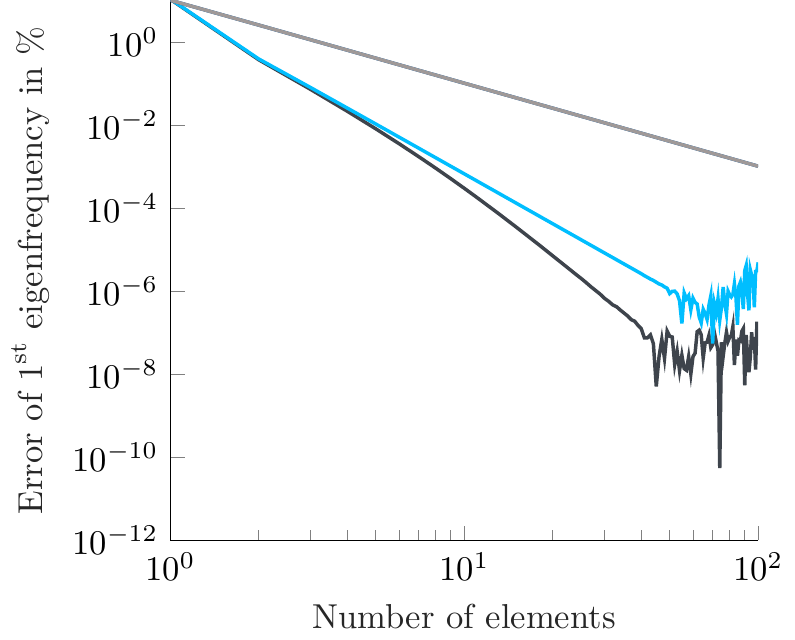} 
	\hfill
	\subcaption{$N_p = N_q = 2$ for axial loads and $N_p = N_q = 4$ for bending.}
\end{subfigure}
\begin{subfigure}[r]{0.5\textwidth}
	\centering
	\includegraphics[width=0.95\textwidth, height=6.5cm]{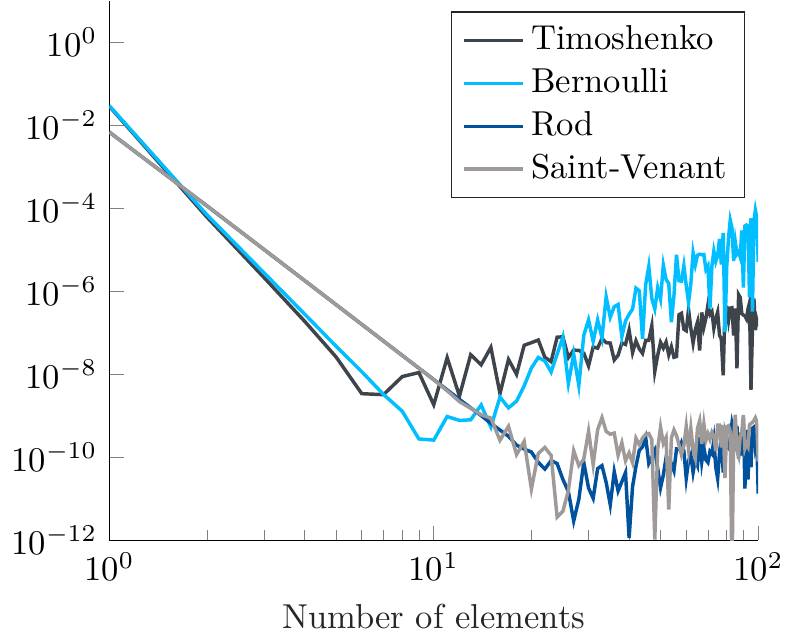} 
	\hfill
	\subcaption{$N_p = N_q = 4$ for axial loads and $N_p = N_q = 6$ for bending.}
\end{subfigure}
\caption[Convergence of PFEM]{Analysis of the approximation error of PFEM for the beam models for a different number of elements and supporting points.}%
\label{fig:convergence}%
\end{figure}

For each of the systems presented in Sec.\,\ref{sec:truss_elements}, a numerical analysis of the approximation error is performed for the case of a single beam with a quadratic cross section and $L/h = 50$, where $h$ is the cross sectional height. 
Different boundary conditions are chosen for bending and axial loads. In case of the Timoshenko and Euler-Bernoulli equations, the beam is simply supported and otherwise it is clamped at one end and free at the other. 
A Poisson ratio of $\nu = 0.1$ is chosen and the Timoshenko shear coefficient is set to $\kappa = 5/6$. 
The beam is then approximated using an increasing number of elements and the error in the first eigenfrequency is computed. 
Analytical solutions for the eigenfrequencies of each of the considered load cases can e.\,g. obtained from \cite{schnell2006technische}. 
The results are depicted in Fig.\,\ref{fig:convergence} for all of the beam equations with the number of elements ranging from $N_\text e = 1 \dots 100$.

In Fig.\,\ref{fig:convergence}~a), $N_p = N_q = 2$ supporting points were chosen in case of the rod and the Saint-Venant element and $N_p = N_q = 4$ for both Euler-Bernoulli and Timoshenko beam. For all systems, the error starts at a value of $\approx 10\,\%$ and decreases monotonically until $N_\text e \approx 40$ where it starts to oscillate around a more or less stationary value for the bending elements. The results for the rod and Saint-Venant element are not discernible from each other. This was to be expected as systems C and D in Sec.\,\ref{sec:truss_elements} only differ in terms of parameters but are otherwise identical. A saturation of the error is not visible for the axial loads. The oscillating behavior about an error of $\approx 1\times 10^{-6} \,\%$ for the Euler-Bernoulli beam and $\approx 5 \times 10^{-8}\,\%$ for the Timoshenko beam is explained by the fact that the composite system becomes numerically ill-conditioned for a high number of elements. This effect can be reduced by stabilizing the numerical methods involved in the system assembly and eigenfrequency calculation.

As the approximation error does not only depend on the number of elements but also on the order of the approximation polynomials, an additional analysis was conducted with a higher number of supporting points. This time, $N_p = N_q = 4$ supporting points were chosen for the axial loads and $N_p = N_q = 6$ for bending. The results are shown in Fig.\,\ref{fig:convergence}~b). With only a single element, the error is now about three orders of magnitudes lower than before for all systems. It also decreases more rapidly and is already below the minimum value visible in Fig.\,\ref{fig:convergence}~a) for $N_\text e = 5$ elements (with the exception of the Timoshenko beam, for which a lower value results at $N_\text e = 74$). However, for $N_e > 10$ the systems start to become numerically ill-conditioned, which results in a saturation of the approximation error for the rod and Saint-Venant elements. Instead of a saturation, an increasing trend is visible for the error of the simply supported Timoshenko and Euler-Bernoulli beams. 

The numerical analysis of the approximation error is limited by the accuracy of the employed numerical methods and the numerical conditioning of the systems for which the error is calculated. A mathematical analysis of the convergence properties of PFEM for the systems considered in this publication would alleviate those issues but is beyond the scope of this publication. With the results shown in Fig.\,\ref{fig:convergence}, we conclude, that reasonable errors can be achieved for a low number of approximating elements, especially when using more than the minimum number of supporting points required per element. However, numerical ill-conditioning becomes a problem for a high number of elements which, depending on the application, needs to properly addressed. Optimizing the algorithms for system assembly on this account is to be considered in further work.

\subsection{High-Rise Building}\label{subsec:demonstrator}

	\begin{figure}
	\begin{subfigure}[l]{0.5\textwidth} 
		\centering
		\includegraphics{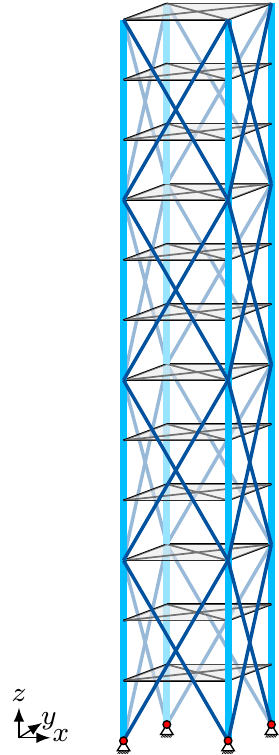} 
		\hfill
		\subcaption{High-rise structure}
	\end{subfigure}
	\begin{subfigure}[r]{0.5\textwidth}
		\centering
		\def\svgwidth{\textwidth}
		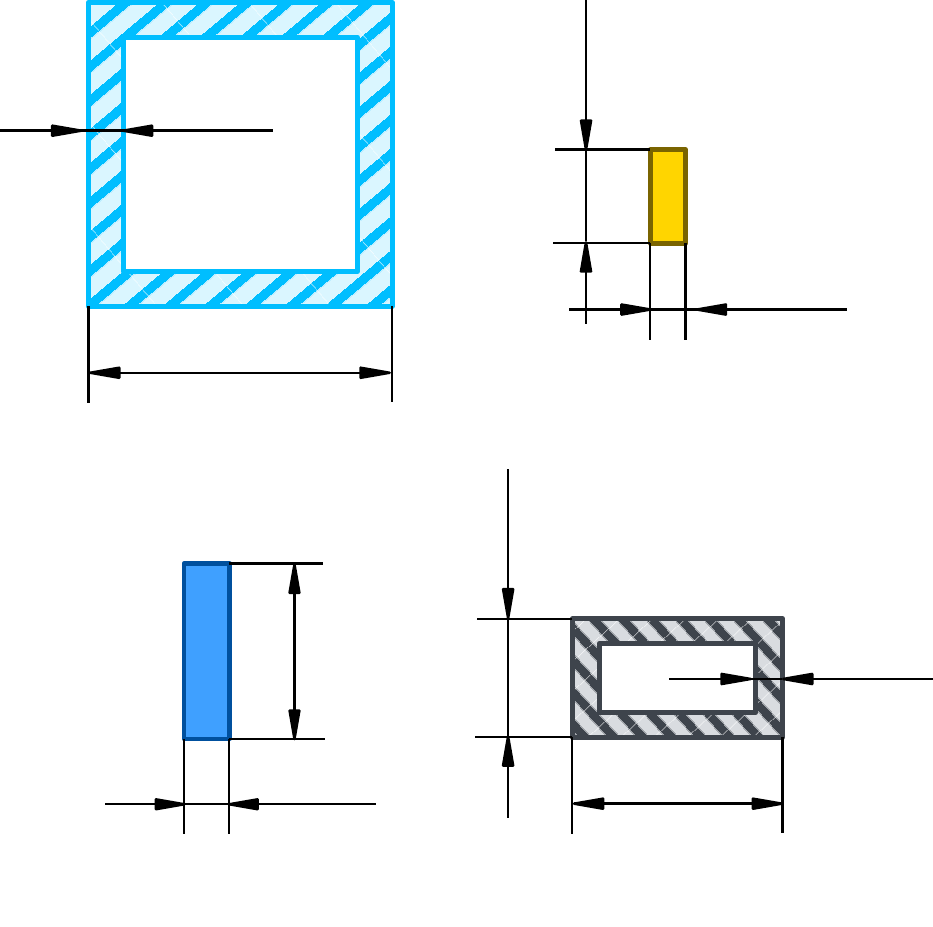
		\subcaption{Beam and rod element cross sections\hfill}
	\end{subfigure}
	\caption{12 story high-rise building composed of beams (columns) and rod elements (all others) including a depiction of the beam and rod element cross sections.}%
	\label{fig:demonstrator}%
	\end{figure}
With the model components and methods described in Secs.\,\ref{sec:truss_elements}-\ref{sec:assembly}, a port-Hamiltonian system model for arbitrary linear elastic 3D truss structures and frames can be built. To facilitate the application of the procedures described in this article, they were compiled into a Matlab framework which is available on GitHub\footnote{\url{http://github.com/awarsewa/ph_fem/}}. 
	
In this section we illustrate the creation of a port-Hamiltonian model for a 12 story high-rise structure. The structure shown in Fig.\,\ref{fig:demonstrator}\,a) is entirely composed of steel bars, $36\,\mathrm{m}$ tall and has a $4.75 \times 4.75\,\mathrm{m}$ footprint. Floor plates are represented by diagonal horizontal bars. The cross sections of columns, horizontal diagonals, horizontal bars and diagonal bracings are depicted in Fig.\,\ref{fig:demonstrator}\,b). Diagonals and horizontal bars are modeled as rod elements with $N_p = N_q = 2$ and can only take axial loads. Columns are composed of two Euler-Bernoulli elements -- one for each bending axis -- with $N_p = N_q = 4$, a rod element and a torsion element with $N_p = N_q = 2$ each. We use Lagrange polynomials for $\boldsymbol \Phi_p$ and $\boldsymbol \Phi_q$ and Gauss-Legendre quadrature for evaluating the integrals in Sec.\,\ref{sec:discretization}. The structure is composed of 4 identical modules. Each of them is composed of 12 columns of length $L = 3\,\mathrm{m}$, 12 horizontal bars, 8 diagonal bracings and 6 horizontal diagonals. This accounts for $n = 392$ states per module and $n = 1568$ states in total with $n_\mathrm{DOF} = 312$ DOFs after concatenation of system matrices according to \eqref{eq:sep}.

Given the node and element table of the structure, the method described in Sec.\,\ref{subsec:const} is used to automatically formulate coupling constraints. All translational DOFs of the lowermost 4 nodes in including the rotation about the $z$-axis are locked. This is achieved by setting the corresponding velocity outputs to zero in \eqref{eq:vel_con}. In total, $n_c = 488$ constraints are formulated for the sample structure in the DAE form \eqref{eq:DAE}. After eliminating all algebraic constraints using the procedure in Sec.\,\ref{subsec:elimination}, the system is reduced to $n = 1080$ ODEs. New system inputs are the external forces $\vec f_\text{ext} \in \mathbb R^{296}$ acting on the structure's DOFs with the collocated outputs being the global nodal velocities $\dot{\vec q}$. Proceeding by elimination of linearly dependent states according to the method in Sec.\,\ref{subsec:linear_dependency}, the number of ODEs is further reduced to $n = 592$, which equals twice the number of DOFs. Then, the transformations from Sec.\,\ref{subsec:to_global} are applied to the system equations to obtain the mass and stiffness matrices $\vec M_\text{fe}$ and $\vec K_\text{fe}$. A Rayleigh-type damping term $\vec D_\text{fe} = \alpha_1 \vec M_\text{fe} + \alpha_2 \vec K_\text{fe}$ is added as in \eqref{eq:ph_iso_mech}, with the damping coefficients set to $\alpha_1 = 0.05$ and $\alpha_2 = 0.005$. 

\begin{figure}
\begin{subfigure}[l]{0.5\textwidth} 
	\centering
	\includegraphics[width=0.9\textwidth, height=4.5cm]{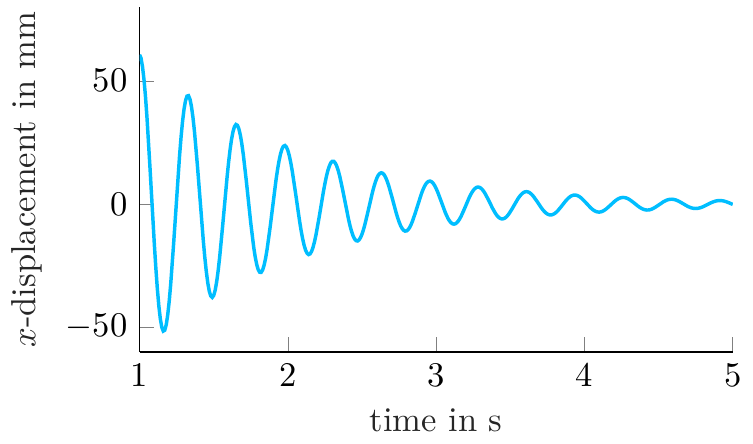} 
	\hfill
	\subcaption{$x$-displacement of node 52\hfill}
\end{subfigure}
\begin{subfigure}[r]{0.5\textwidth}
	\centering
	\includegraphics[width=0.9\textwidth, height=4.5cm]{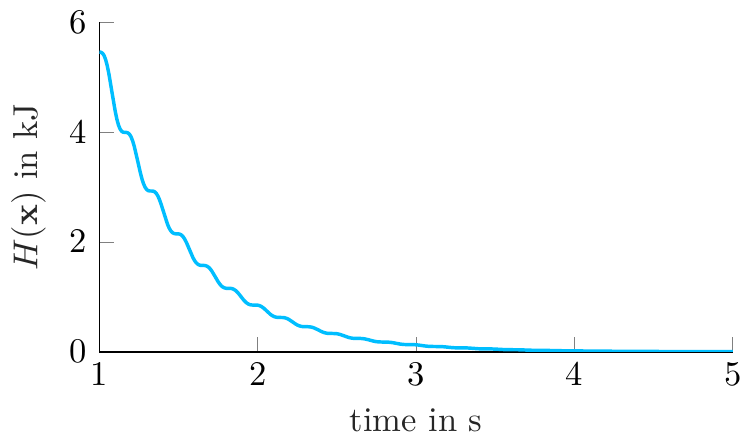} 
	\subcaption{System Hamiltonian\hfill}
\end{subfigure}
\caption{Structural response of the structure shown in Fig.\,\ref{fig:demonstrator}\,a) for an initial displacement in $x$-direction.}%
\label{fig:initial_displacement}%
\end{figure}

\begin{table}
	\caption{First six eigenfrequencies of the structure in Fig.\,\ref{fig:demonstrator}\,a).}
	\label{tab:eigenfrequencies}
	\centering
	\begin{tabular}{lcccccc}
		Eigenmode \#	& 1 & 2 & 3 & 4 & 5 & 6 \\ \hline
		port-Hamiltonian & 3.0699\,Hz & 3.0700\,Hz & 4.5339\,Hz & 8.7633\,Hz & 8.7635\,Hz & 9.1723\,Hz \\
		conventional FE & 3.0700\,Hz & 3.0700\,Hz & 4.5337\,Hz & 8.7632\,Hz & 8.7632\,Hz & 9.1725\,Hz \\
	\end{tabular}
\end{table}

We simulated the dynamic response of the structure for an initial displacement $\vec x_0$, which was calculated by assuming a stationary wind load acting in the positive $x$-direction. The wind load increases linearly with the story number and amounts to $f_\text{wind} = 20\,\mathrm{kN}$ for the uppermost left nodes. The structural displacement of node 52 (top of the structure) is shown in Fig.\,\ref{fig:initial_displacement}\,a). Oscillation caused by the initial displacement is damped almost completely within five seconds. This is also reflected in the system energy shown in terms of the Hamiltonian $H(\vec x)$ in Fig.\,\ref{fig:initial_displacement}\,b).

For comparison, $\vec M_\text{fe}$ and $\vec K_\text{fe}$ were assembled according to a conventional FE method by adding up the element mass and stiffness matrices after transforming them to global coordinates. Element mass and stiffness matrices for the Euler-Bernoulli beam can be found e.\,g. in \cite{schwertassek2017dynamik}. The structural response of the system obtained the conventional way, when simulating it with the same initial condition, is almost identical to the one shown in Fig.\,\ref{fig:initial_displacement}. There are minor differences caused by the respective numerical methods involved in the calculation of $\vec M_\text{fe}$ and $\vec K_\text{fe}$. 

A comparison between the first six eigenfrequencies of the structure obtained by the method presented in this article and by the conventional FE approach is shown in Tab.\,\ref{tab:eigenfrequencies}. Except for the second eigenmode, the eigenmodes deviate slightly in the fourth decimal place. The first and second, as well as the fourth and fifth modes, are bending modes in the $x$- and $y$-direction. Due to the symmetry of the structure, their eigenfrequencies should be identical, as observed in case of the conventional FE approach. The fact that they differ slightly for the presented approach is again explained by the numerics involved. Mass and stiffness matrix are obtained after a number of transformations, while analytical formulations of the approximated element mass and stiffness matrices are used in conventional FEM. The only transformation involved in assembling the global mass and stiffness matrix in the latter case is the rotation of local coordinate frames which makes the method less susceptible to numerical errors. Thus, the higher number of transformations and numerical methods involved in the presented framework can be viewed as a disadvantage. Since it is not strictly necessary, the transformation to global coordinates explained in Sec.\,\ref{subsec:to_global} can be omitted to reduce numerical error. Especially when the mechanical parts of the system are coupled with non-mechanical or hybrid components.  An example of such a multi-domain system is presented in the following section.

\subsection{Multi-Domain System}
In a recent publication of some of the authors, a method for the integration of actuators into FE models of adaptive structures is presented\,\cite{boehm2020input}. To allow for an easy integration of active elements, a quasi-stationary assumption is made for the local dynamics of passive elements. While this is a valid simplification for most large structures, the physical coupling between the dynamics of the actuators and those of the structure is lost this way. If this is not desired, a significant increase in complexity results. In this section, we show that it is straightforward to integrate hydraulic actuators into a mechanical structure within the presented port-Hamiltonian modeling framework. Simplifications such as the ones in\,\cite{boehm2020input} need not be made.

\begin{figure}
	\begin{subfigure}[l]{0.45\textwidth} 
		\centering
		\def\svgwidth{\textwidth}
		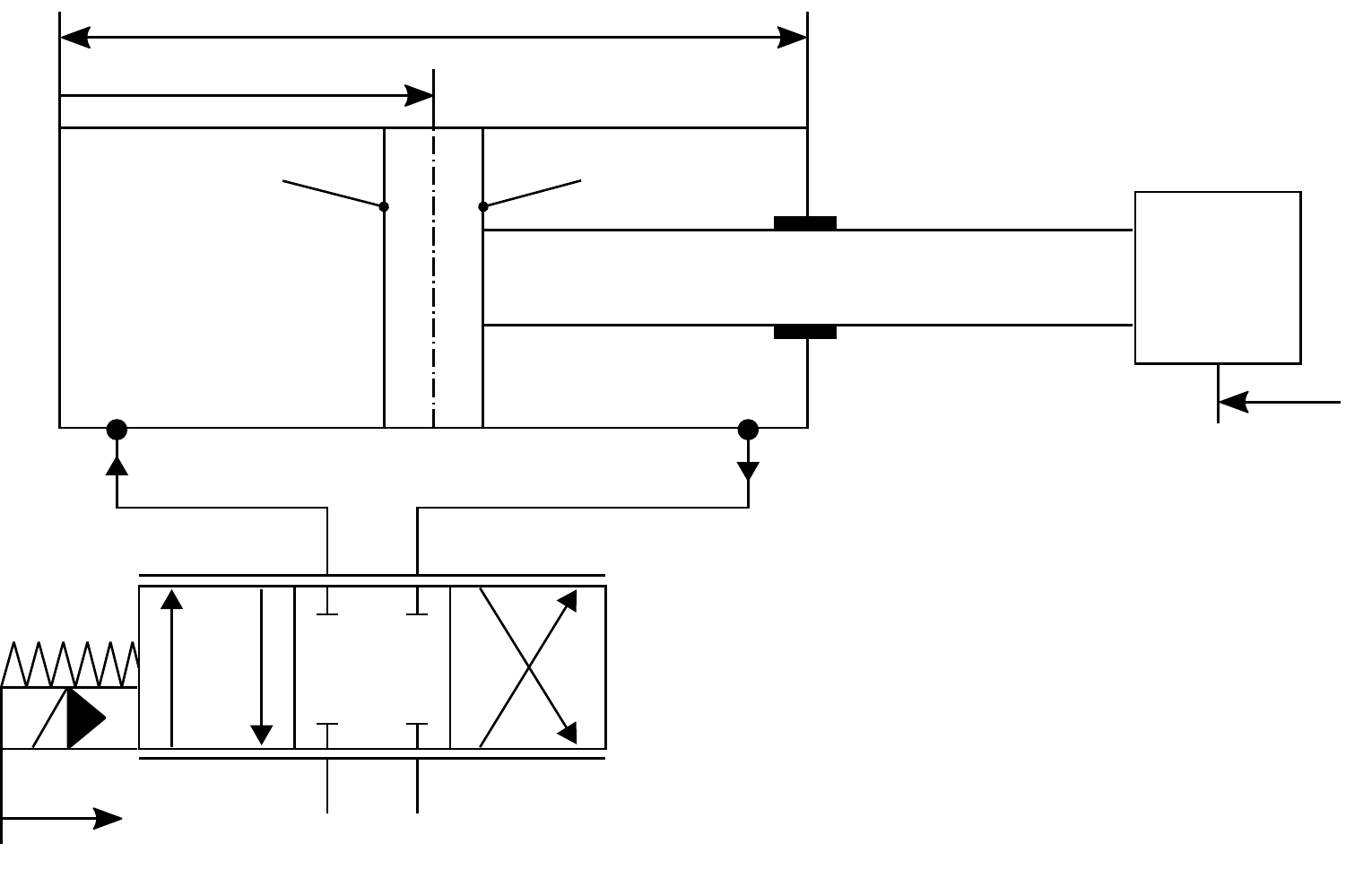
		\hfill
		\subcaption{Double-acting hydraulic cylinder with valve\hfill}
	\end{subfigure}
	\begin{subfigure}[r]{0.55\textwidth}
		\centering
		\includegraphics{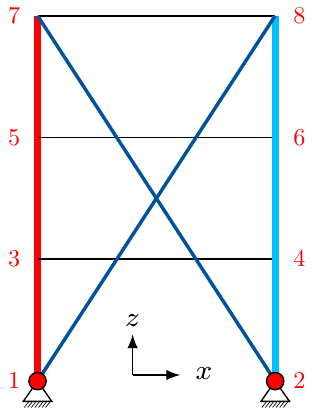} 
		\hfill
		\subcaption{Steel frame with parallel hydraulic cylinder (red)}
	\end{subfigure}
	\caption{Multi-Domain model of an adaptive structure that is actuated by a hydraulic cylinder.}%
	\label{fig:frame_with_hydraulics}%
\end{figure}

For the coupling of a mechanical structure with a hydraulic actuator, a port-Hamiltonian model of the latter is required. We use a slightly modified version of the model of a valve-controlled double-acting hydraulic cylinder presented by Kugi and Kemmetmüller in\,\cite{kugi2004new} for which we will briefly present the necessary equations in the following. For details, the reader is referred to\,\cite{kugi2004new}. A schematic representation of the piston actuator with a three-land-four-way valve is shown in Fig.\,\ref{fig:frame_with_hydraulics}~a). The volumetric flows into the cylinder chambers are given by
\begin{equation}
\dot q_1 = \Gamma_1 x_\text v = \begin{cases} k_\text v \sqrt{p_\text S - p_1} x_\text v & \text{for}~x_\text v \geq 0, \\ k_\text v \sqrt{p_1 - p_\text T} x_\text v & \text{for}~x_\text v < 0 \end{cases}, \qquad \dot q_2 = \Gamma_2 x_\text v = \begin{cases} k_\text v \sqrt{p_2 - p_\text T} x_\text v & \text{for}~x_\text v \geq 0, \\ k_\text v \sqrt{p_\text S - p_2} x_\text v & \text{for}~x_\text v < 0 \end{cases}
\end{equation}
where $x_\text v$ and $k_\text v$ are the valve displacement and the valve coefficient, $p_\text S$ and $p_\text T$ are the supply and the tank pressure and $p_1$ and $p_2$ the respective chamber pressures. With the common simplification of a linearized constitutive law of the bulk modulus $\beta$, the system dynamics of the hydraulic cylinder can be formulated as follows
\begin{gather}
	\dot s = \frac{1}{m} w_s, \quad \dot w_s = p_1 A_1 - p_2 A_2 - F_\text{ext}, \label{eq:force_balance_piston} \\
	\dot p_1 = \frac{\beta}{A_1 s} (-A_1 w_s + \Gamma_1 x_\text v), \quad \dot p_2 = -\frac{\beta}{A_2 (L-s)}(A_2 w_s - \Gamma_2 x_\text v).
\end{gather}
Here, $s$ is the piston position, $m$ its mass and $w_s$ its impulse (in\,\cite{kugi2004new} the velocity). Multiplying the chamber pressures with the effective piston areas $A_1$ and $A_2$ and taking into account the external force $F_\text{ext}$ yields the force balance equation in \eqref{eq:force_balance_piston}. For the system Hamiltonian we get
\begin{equation}
H = U + \frac{1}{2m} w_s^2, \quad \text{with } U = A_1 s\left(\beta \left(\exp\left(\frac{p_1}{\beta}\right)-1\right)-p_1\right) + A_2 (L-s) \left( \beta \left(\exp\left(\frac{p_2}{\beta}\right) - 1\right) -p_2 \right).
\end{equation}
See\,\cite{kugi2004new} or\,\cite{kugi2003energy} for the derivation of the internal energy $U$. Everything necessary for the formulation of the actuator dynamics as a nonlinear port-Hamiltonian system has now been defined. Choosing the state vector as $\vec x = \begin{bmatrix} s & w_s & p_1 & p_2 \end{bmatrix}^\text T$ and the input as $\vec u = \begin{bmatrix} x_\text v & F_\text{ext} \end{bmatrix}^\text T$, we get
\begin{align}
	\begin{split}
		\dot{\vec x} &= \vec J(\vec x) \frac{\partial H}{\partial \vec x}(\vec x) + \vec g(\vec x) \vec u, \\
		\vec y &= \vec g^\text T(\vec x) \frac{\partial H}{\partial \vec x}(\vec x),
	\end{split}
\end{align}
where
\begin{equation}
\vec J(\vec x) = \begin{bmatrix} 0 & 1 & 0 & 0 \\ -1 & 0 & \frac{\beta}{s} & -\frac{\beta}{L-s} \\
0 & -\frac{\beta}{s} & 0 & 0 \\ 0 & \frac{\beta}{L-s} & 0 & 0 \end{bmatrix}, \quad \vec g(\vec x) = \begin{bmatrix} 0 & 0 \\ 0 & -1 \\ \frac{\beta}{A_1 s}\Gamma_1 & 0 \\ \frac{\beta}{A_2(L-s)} \Gamma_2 & 0 \end{bmatrix}.
\end{equation}
The hydraulic cylinder is connected to the steel frame shown in Fig.\,\ref{fig:frame_with_hydraulics}~b) in parallel to the left vertical column. To simplify things a little, the frame is taken to be one side of the lowermost module (three stories) of the high-rise building shown in Fig.\,\ref{fig:demonstrator}~a). It is composed of the same elements and the same parameters as in Sec.\,\ref{subsec:demonstrator} are used for both approximation and damping. 
Coupling of the frame and the cylinder can be done just like before using the procedure described in Sec.\,\ref{sec:assembly} until a system of the form \eqref{eq:DAE} is obtained. Since the coupled system is now nonlinear, we cannot proceed with the elimination of algebraic constraints in an automated fashion. However, it is still possible to obtain an explicit formulation by solving for the Lagrange multipliers (see e.\,g. \cite{duindam2009modeling} and \cite{van2014port})
\begin{equation}
	\boldsymbol \lambda(\vec x, \vec u) = -\left( \vec B^\text T(\vec x)\frac{\partial^2 H}{\partial \vec x^2}(\vec x) \vec B(\vec x) \right)^{-1}\vec B(\vec x) \frac{\partial^2 H}{\partial \vec x^2}(\vec x) \left(\vec J(\vec x)\frac{\partial H}{\partial \vec x}(\vec x) + \vec g(\vec x) \vec u \right).
\end{equation}

\begin{figure}
\begin{subfigure}[l]{0.45\textwidth} 
	\centering
	\includegraphics[width=0.95\textwidth, height=5.5cm]{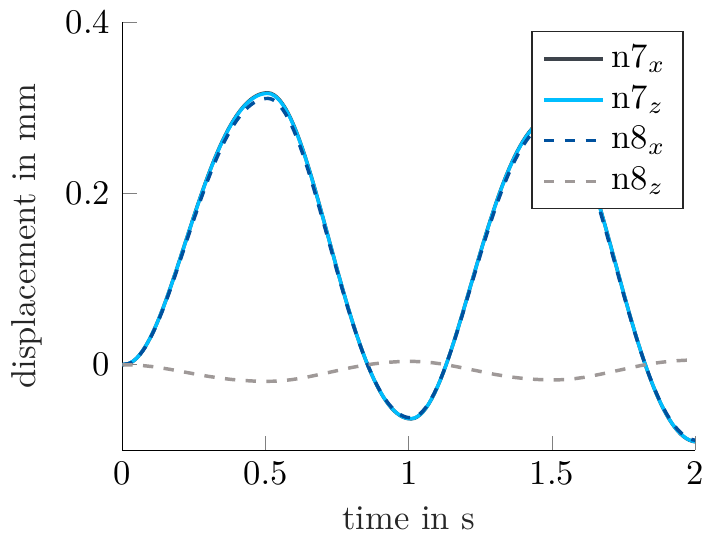} 
	\hfill
	\subcaption{Displacements of the frame's topmost nodes}
\end{subfigure}
\begin{subfigure}[r]{0.55\textwidth}
	\centering
	\includegraphics[width=0.95\textwidth, height=5.5cm]{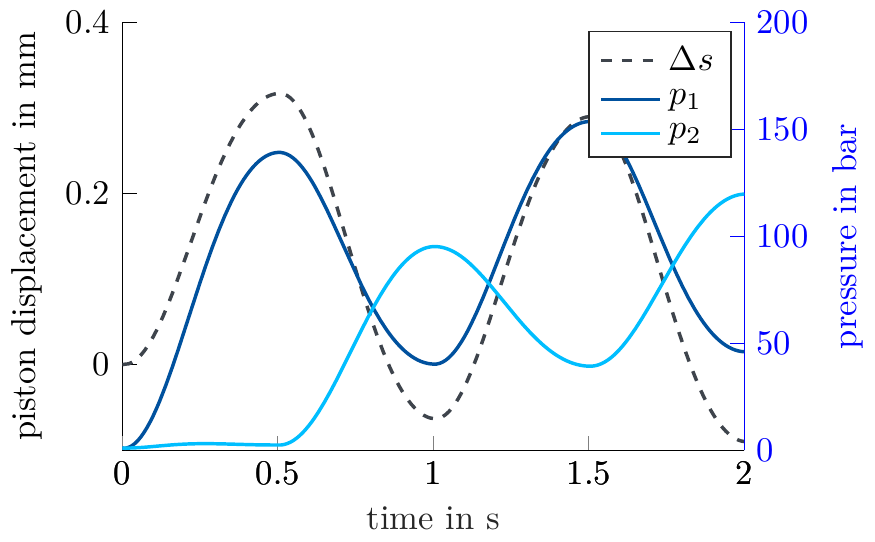} 
	\hfill
	\subcaption{Hydraulic cylinder dynamics}
\end{subfigure}
\caption{Structural response of the system shown in Fig.\,\ref{fig:frame_with_hydraulics} for sinusoidal excitation of the hydraulic cylinder}%
\label{fig:res_hydraulics}%
\end{figure}

A simulation of the coupled multi-domain system is shown in Fig.\,\ref{fig:res_hydraulics}, where the valve displacement $x_\text v$ follows a sinusoidal excitation. For the cylinder, the length was set to $L = 35\,\mathrm{cm}$, the effective piston areas were chosen as $A_1 = 133\,\mathrm{cm}^2$ and $A_2 = 94.2\,\mathrm{cm}^2$ and the piston mass as $m = 10\,\mathrm{kg}$. The system is driven with a supply pressure of $p_\text S = 200\,\mathrm{bar}$ and a tank pressure of $p_\text T = 0\,\mathrm{bar}$. The initial pressures in the chambers were set to $p_1 = p_2 = 1\,\mathrm{bar}$ and the initial piston position was set to $s = 10\,\mathrm{cm}$. A bulk modulus of $\beta = 1.3\,\mathrm{GPa}$ and a valve coefficient of $x_\text v = 2\,\mathrm{l/min}$ were assumed.

In Fig.\,\ref{fig:res_hydraulics}~a), the $x$- and $z$-displacements of the topmost nodes of the frame are shown, whereas Fig.\,\ref{fig:res_hydraulics}~b) depicts the chamber pressures and the piston displacement $\Delta s$. The coupling between piston displacement and the motion of the frame is immediately visible. Comparing the $z$-displacement of node 7 in Fig.\,\ref{fig:frame_with_hydraulics}~a) with $\Delta s$ in Fig.\,\ref{fig:frame_with_hydraulics}~b), especially at the peaks at $t = 0.5\,\mathrm{s}$ and $t = 1.0\,\mathrm{s}$, shows that they are identical. 
If the frame dynamics are modeled with a standard FEM approach and the hydraulic cylinder separately in the hydraulic domain, physical coupling of the two systems is not as straightforward. In the port-Hamiltonian modeling approach, coupling via the power flow at the ports using inputs and outputs is a core procedure and can be handled without difficulties. 
With the example, we demonstrate the feasibility of the presented approach to model multi-domain systems and also that it is not limited to linear systems.

\section{Conclusion and Further Work}\label{sec:conclusion}
In this contribution, a selection of methods for the modeling of the dynamics of complex structures as linear ISO port-Hamiltonian systems was presented. Starting from the distributed parameter formulations of axial, torsional and bending motion of a beam element, the corresponding infinite-dimensional port-Hamiltonian systems were derived. In order to enable numerical simulations of these systems, finite-dimensional approximations are necessary. Those were obtained via a structure preserving mixed finite element method. A numerical analysis of the approximation error was presented for the beam equations. \color {black}Given the spatially discretized models for individual load cases, complex structures can be assembled by interconnecting them via coupling constraints. 
This leads to index one DAEs that can be reduced to an ODE system on a constraint state space by means of a transformation. 
Afterward, further reduction of the system order is achieved by eliminating linearly dependent effort variables. 
The resulting system coordinates can be transformed to the space of global DOFs, which allows comparison of the system's mass and stiffness matrices to the ones obtained using conventional FE methods. The process was presented for a multi-story high-rise building. 

When a system is only composed of mechanical components, the mass and stiffness matrices from a conventional FE model can also be used to formulate a port-Hamiltonian system. In fact, this is advantageous, since the number of transformations and numerical routines involved in the presented approach is higher, which results in a higher numerical error. If the aim is, however, to model heterogeneous multi-domain systems as port-Hamiltonian systems, the merit of the presented method is, that non-mechanical elements can be included in each step of the process. Since the port-Hamiltonian structure is preserved in each step, this makes it easy to add such elements on both the PDE and ODE level.
We demonstrated the coupling of the hydraulic with the mechanical domain which is straightforward with the presented methods. It also shows that the methodology is not limited to linear systems.

A Matlab framework which encompasses all the models and methods presented in this contribution is available online. Not only can it be used to reproduce the results in Sec.\,\ref{sec:example}, but also to build arbitrary structures composed of the elements introduced in Sec.\,\ref{sec:truss_elements}. The framework is written in an object-oriented way, such that the user only needs to provide the geometry and material parameters of the elements including a node and element table. Different examples are provided which illustrate the usage of the methods. Extension of the code to account for e.\,g. different types of damping or model order reduction is a matter of ongoing work. Also an optimization with respect to numerical stability and precision.

Currently, coupling of simple elements is based on a block diagram analogy. In further work, we aim to represent component interconnection in bond graphs which is more in line with port-Hamiltonian systems theory and facilitates interfacing with non-mechanical components and comprehension of system characteristics. To model a wider range of adaptive structures, further subsystems such as plate and shell elements or integrated fluidic actuators need to be included in the framework. Furthermore, we also want to develop control and state estimation methods based on the models presented in this article.

\section*{Acknowledgment}
The authors gratefully acknowledge the generous funding of this work by the German Research Foundation (DFG - Deutsche Forschungsgemeinschaft) as part of the Collaborative Research Center 1244 (SFB) "Adaptive Skins and Structures for the Built Environment of Tomorrow"/projects A06 and B02.

\appendix
\section{Intermediate Steps in Discretization}\label{app:approx}
Since it is not immediately clear how to obtain \eqref{eq:partial} from \eqref{eq:weak_system} using integration by parts, some intermediate steps are shown here to enhance comprehensibility. We begin with the term including the first order spatial derivative for which we obtain
\begin{align}\label{eq:A1}
		\int_Z \boldsymbol \phi_p A_1 \frac{\partial}{\partial z} \boldsymbol \phi_q^\mathrm{T} \vec e_q \,\mathrm{d}z &= - \int_Z \frac{\partial}{\partial z} \boldsymbol \phi_p A_1 \boldsymbol \phi_q^\mathrm{T} \vec e_q \,\mathrm{d}z + \left[ \boldsymbol \phi_p A_1 \boldsymbol \phi_q^\mathrm{T} \vec e_q \right]_a^b 
\end{align}
and are finished in case our system has no higher-order spatial derivatives. For the second order spatial derivative, integration by parts needs to be applied twice. The term 
\begin{equation}
	\int_Z \boldsymbol \phi_p A_2 \frac{\partial^2}{\partial z^2} \boldsymbol \phi_q^\mathrm{T} \vec e_q \,\mathrm{d}z 
\end{equation}
is omitted on the left hand side in the following due to limited space and we proceed with
\begin{align}\label{eq:A3}
	 \begin{split}
		\dots &= - \int_Z \frac{\partial}{\partial z} \boldsymbol \phi_p A_2 \frac{\partial}{\partial z} \boldsymbol \phi_q^\mathrm{T} \vec e_q \,\mathrm{d}z + \left[ \boldsymbol \phi_p A_1 \frac{\partial}{\partial z} \boldsymbol \phi_q^\mathrm{T} \vec e_q \right]_a^b \\
		&= \int_Z \frac{\partial^2}{\partial z^2} \boldsymbol \phi_p A_2 \boldsymbol \phi_q^\mathrm{T} \vec e_q \,\mathrm{d}z + \left[ \boldsymbol \phi_p A_2 \frac{\partial}{\partial z} \boldsymbol \phi_q^\mathrm{T} \vec e_q \right]_a^b - \left[ \frac{\partial}{\partial z} \boldsymbol \phi_p A_2 \boldsymbol \phi_q^\mathrm{T} \vec e_q \right]_a^b.
		\end{split} 
\end{align}
Accordingly, integration by parts needs to be repeated N times to 
\begin{equation} 
	\int_Z \boldsymbol \phi_p A_N\frac{\partial^N}{\partial z^N} \boldsymbol \phi_q^\mathrm{T} \vec e_q \,\mathrm{d}z 
\end{equation}
which yields
\begin{align}\label{eq:A5}
	 \begin{split}
		\dots =& - \int_Z \frac{\partial}{\partial z} \boldsymbol \phi_p A_N \frac{\partial^{N-1}}{\partial z^{N-1}} \boldsymbol \phi_q^\mathrm{T} \vec e_q \,\mathrm{d}z + \left[ \boldsymbol \phi_p A_N \frac{\partial^{N-1}}{\partial z^{N-1}} \boldsymbol \phi_q^\mathrm{T} \vec e_q \right]_a^b \\
		=& \dots \\
		=& (-1)^{N} \int_Z \frac{\partial^N}{\partial z^N} \boldsymbol \phi_p A_N \boldsymbol \phi_q^\mathrm{T} \vec e_q \,\mathrm{d}z + \left[ \boldsymbol \phi_p A_N \frac{\partial^{N-1}}{\partial z^{N-1}} \boldsymbol \phi_q^\mathrm{T} \vec e_q \right]_a^b - \dots \\
		& + (-1)^{N-1} \left[ \frac{\partial^{N-1}}{\partial z^{N-1}} \boldsymbol \phi_p A_N \boldsymbol \phi_q^\mathrm{T} \vec e_q \right]_a^b.
		\end{split} 
\end{align}
Merging \eqref{eq:A1}, \eqref{eq:A3} and \eqref{eq:A5} and combining the bracketed integrals in sums respectively, it can be seen that the result of applying integration by parts $N$ times is indeed given by \eqref{eq:partial}.

\bibliographystyle{elsarticle-num}

\end{document}